\documentclass{elsart}

\usepackage{amsmath,amssymb,amsfonts}
\usepackage{bm}
\usepackage{graphicx}

\newcommand{\be}{\begin{equation}}
\newcommand{\ee}{\end{equation}}
\newcommand{\ba}{\begin{eqnarray}}
\newcommand{\ea}{\end{eqnarray}}
\newcommand{\bma}{\left(\begin{array}}
\newcommand{\ema}{\end{array}\right)}
\def\bs{\begin{subequations}}
\def\es{\end{subequations}}
\def\a{\alpha}
\def\b{\beta}
\def\g{\gamma}
\def\e{\epsilon}
\def\t{\theta}
\def\s{\sigma}
\def\cE{{\cal E}}

\def\p{\partial}
\newcommand{\Eq}[1]{(\ref{#1})}
\def\lp{\ell_{\rm Pl}}
\def\rmi{{\rm i}}
\def\rme{{\rm e}}
\def\rmd{{\rm d}}

\begin{document}
\begin{frontmatter}

\title{Quantum gravity as a Fermi liquid}
\author{Stephon H.S. Alexander},
\ead{stephonalexander@mac.com}
\address{Institute for Gravitation and the Cosmos, Department of Physics,\\ The Pennsylvania State University,
104 Davey Lab, University Park, PA 16802}
\address{Department of Physics, Haverford College, 370 Lancaster Avenue, Haverford, PA 19041}
\author{Gianluca Calcagni}
\ead{gianluca@gravity.psu.edu}
\address{Institute for Gravitation and the Cosmos, Department of Physics,\\ The Pennsylvania State University,
104 Davey Lab, University Park, PA 16802}
\date{\today}
\begin{abstract}
We present a reformulation of loop quantum gravity with a cosmological constant and no matter as a Fermi-liquid theory. When the topological sector is deformed and large gauge symmetry is broken, we show that the Chern--Simons state reduces to Jacobson's degenerate sector describing $1+1$ dimensional propagating fermions with nonlocal interactions.  The Hamiltonian admits a dual description which we realize in the simple BCS model of superconductivity. On one hand, Cooper pairs are interpreted as wormhole correlations at the de Sitter horizon; their number yields the de Sitter entropy. On the other hand, BCS is mapped into a deformed conformal field theory reproducing the structure of quantum spin networks. When area measurements are performed, Cooper-pair insertions are activated on those edges of the spin network intersecting the given area, thus providing a description of quantum measurements in terms of excitations of a Fermi sea to superconducting levels. The cosmological constant problem is naturally addressed as a nonperturbative mass-gap effect of the true Fermi-liquid vacuum.
\end{abstract}

\begin{keyword}
Loop quantum gravity \sep Cosmological constant \sep Fermi-liquid theories
\PACS 04.60.Pp, 74.20.-z
\end{keyword}
\end{frontmatter}


\section{Introduction}

Background independence is a property often translated in the requirement of the metric to be as dynamical as all other matter fields. Loop quantum gravity (LQG) is an example of how to perform a background independent quantization of general relativity \cite{rov97,thi02,smo04,rov04}. One of the challenges of the LQG program is to find a ground state that gives a semiclassical description of the observed universe; the cosmological concordance model points to a vacuum which is asymptotically de Sitter (dS). In the absence of matter fields, a candidate ground state annihilates the Hamiltonian constraint $\hat {\cal H}|\Psi\rangle=0$ (Wheeler--deWitt equation in the cosmological case \cite{ACS}). In quantum field theory the vacuum corresponds to a state with zero occupation number, so that matter creation operators will excite particles from $|\Psi\rangle$. It is well known that such a vacuum state is not diffeomorphism invariant \cite{BD}. So a question to ask is whether loop quantum gravity admits a vacuum state that simultaneously yields a vanishing expectation value for the gravitational Hamiltonian and annihilates matter fields. We shall argue in favour of a positive answer which is relevant also as regards the cosmological constant problem. In fact, the coupling of matter to geometry is intrinsically encoded in the quantum geometry rather than represented as a fluctuation on a fixed background. We show how the holomorphic ground state wavefunction of quantum gravity can naturally encode fermionic degrees of freedom without resort to a background.\footnote{At the classical level one indeed selects a particular background, namely de Sitter. Nonetheless, background independence is realized in LQG by considering superpositions of geometric states.}

Canonical general relativity can be completely expressed in a manner giving it close semblance to a first-order nonabelian gauge theory. This is accomplished with the Ashtekar connection \cite{ash86}, an $SU(2)$ complex-valued gauge field. This recasting allows for a nonperturbative quantization of gravity. Matter can just be added into by hand without changing the structure of the quantum theory. However, an ambiguity in the geometric operators (the Immirzi parameter) cannot get decoupled when fermionic matter is included \cite{PeR}. In \cite{AV1,AV2} it was argued that a gap in the spectrum of fermions can tie in the Immirzi parameter with the cosmological constant so as to relax the cosmological constant problem. There, the fermions were introduced by hand. In this paper it is argued that these fermions emerge naturally from extended, nonlocal flux configurations in a deformed topological sector of LQG. This sector can be thought of as arising from a spontaneous symmetry breaking induced by one-dimensional gravielectric flux configurations which terminate on monopole defects. On one hand, geometry is given an interpretation in terms of a lower-dimensional fermionic theory and monopoles are related to pair-wise interactions in a Fermi liquid. On the other hand, they generate a structure strongly resembling that of framed spin networks. This relationship gives us a holographic description of the entropy which resides on the de Sitter horizon in terms of the quantum dynamics of Cooper pairs (i.e., boundary wormholes) in $1+1$ dimensions.

The main strategy to link all these elements was outlined in \cite{AC}; here it is developed in greater detail along the following steps:
\begin{enumerate}
\item Add a topological term, depending on a parameter $\t$, to the $3+1$ Hamiltonian.
\item Quantize the constraints and apply them on the Chern--Simons (often called Kodama) state.
\item Deform the topological parameter $\t$ to a functional dependent on the connection, $\t\to\t(A)$. Two of the classical $(E,A)$ configurations compatible with this deformation are degenerate. One of them, which will be our subject of interest, is Jacobson's sector. Some of the reasons why to choose this sector are: (i) there are no anomaly issues after the cosmological constant becomes dynamical; (ii) the theory living in it admits a well-defined path-integral quantization in terms of fermionic variables; (iii) properties of the four-dimensional theory have a simple holographic origin from the degenerate model.
\item The latter is a Fermi-liquid theory whose ground state is precisely the deformed Chern--Simons state.
\item As a concrete realization of a quantum spin network with effective cosmological constant we consider the Bardeen--Cooper--Schrieffer (BCS) model, a deformed conformal field theory (CFT) which first successfully described superconductive media.
\end{enumerate}
To deform $\t$ leads to an altogether new theory, which one would like to define already at the level of the total Hamiltonian or action. However, for the time being our goal is to reduce it to ordinary loop quantum gravity when the deformation is very weak (perturbative limit). An expansion of the new ground state and total Hamiltonian in a perturbative parameter must yield, at leading order, just LQG with a cosmological constant. This is sufficient to justify our choice of taking a constraint and a state of the usual theory and deform both into some new constraint annihilating a new state.

Sections \ref{hami} (Ashtekar formalism and topological terms), \ref{csstate} (Hamiltonian quantization and Chern--Simons state), \ref{jaco} (Jacobson's degenerate sector), \ref{bcs} (BCS theory), and \ref{fsn} (framed spin networks) review the needed ingredients in a self-contained way for the reader unfamiliar with either of them.

The main results are presented in Sect.~\ref{main}, where quantum gravity in the degenerate sector is shown to behave as a Fermi liquid, and Sect.~\ref{inte}, where framed spin networks of critical level are interpreted as a BCS liquid living on a two-dimensional network. The main equations are Eqs.~\Eq{mainl1}, \Eq{mainl2}, \Eq{mass2} and \Eq{bj-}. The quantum degenerate sector is also associated with the boundary of a de Sitter background, where there lives a magnetic monopole which is the gravitational analogue of a Yang--Mills soliton. Concluding remarks are in Sect.~\ref{disc}. 


\section{Classical gravity with topological terms}\label{hami}

In the Ashtekar formalism \cite{ash86}, gravitational dynamics on a four-dimensional manifold ${\cal M}_4$ is not described by a metric $g_{\mu\nu}$ but, rather, by the real-valued gravitational field $e_\mu^I(x)$, mapping a vector $v^\mu$ in the tangent space of ${\cal M}_4$ at the point $x$ into Minkowski spacetime $M_4$ (with metric $\eta_{IJ}={\rm diag}(-1,1,1,1)_{IJ}$). Locally, the metric on ${\cal M}_4$ is $g_{\mu\nu}=\eta_{IJ}e_\mu^Ie_\nu^J$. Hereafter, Greek indices $\mu,\nu,\dots$ are spacetime, $\a,\b,\dots$ are spatial, capital Latin letters $I,J,\dots$ denote directions in the internal gauge space while $i,j,\dots$ are restricted to its spatial sections. Upper and lower equal indices are conventionally summed over. A notation often adopted makes use of the exterior algebra in the wedge product $\wedge$, defining the spatial volume form as $\rmd x^\a\wedge \rmd x^\b\wedge \rmd x^\g=\e^{\a\b\g}\rmd^3x$. $\epsilon_{\a\b\g}$ is the Levi--Civita symbol, the structure constant of the $su(2)$ algebra: $\epsilon_{123}=\epsilon_{231}=\epsilon_{312}=+1$, antisymmetric in pairs of indices, indices raised by Kronecker delta. It obeys the identity $\e_{\a\b\g}\e^{\a\b'\g'}=\delta_\b^{\b'}\delta_\g^{\g'}-\delta_\g^{\b'}\delta_\b^{\g'}$.
The Lorentz connection $\omega_{\mu I}^{\phantom{\mu I}J}$ is 
$\omega_{I}^{\phantom{I}J}\equiv\omega_{\mu I}^{\phantom{\mu I}J}\rmd x^\mu$,
$\rmd\omega_I^{\phantom{I}J}\equiv\p_\mu \omega_{\nu I}^{\phantom{\mu I}J}\rmd x^\mu\wedge \rmd x^\nu$ is the exterior derivative, and the curvature of $\omega$ is
$R_I^{\phantom{I}J}=\rmd\omega_I^{\phantom{I}J}+\omega_I^{\phantom{I}K}\wedge\omega_K^{\phantom{I}J}$. The action of self-dual gravity is (up to the gravitational constant $G$)
\be\label{act}
S=\int_{{\cal M}_4} [*(e^I\wedge e^J)\wedge R_{IJ}  + \rmi e^I\wedge e^J\wedge R_{IJ}]\,,
\ee
where $*$ is the Hodge dual, the first term is the Hilbert--Palatini action and the second is the Holst term (proportional to the first Bianchi identities in the absence of torsion).

Here we are interested in the Hamiltonian formulation in Ashtekar variables \cite{ash86,sam87}. In the gauge choice $e^0_\mu=0$, it is convenient to define the densitized triad $E^\a_i\equiv\e_{ijk}\e^{\a\b\g} e^j_\b e^k_\g$, which is conjugate to the self-dual connection
\be
A_\a^i(x)\equiv-\tfrac12 \e^{ij}_{\phantom{ij}k}\omega_{\a j}^{\phantom{\a j}k}-\rmi\omega_{\a 0}^{\phantom{\a 0}i}.
\ee
As the Lorentz connection (and, in particular, the spin connection $\Gamma_\a^i\equiv -\tfrac12 \e^{ij}_{\phantom{ij}k}\omega_{\a j}^{\phantom{\a j}k}$) is real in real gravity, $A$ is complex-valued and obeys the reality conditions (for a discussion, see e.g. \cite{Men93})
\be\label{rc}
A_\a^{i*}+ A_\a^i=2\Gamma_\a^i[E]\,,
\ee
where $*$ denotes complex conjugation and the spin connection solves the equation $\rmd e+\Gamma[E]\wedge e=0$.

The Poisson bracket of the elementary variables $A$ and $E$ is (equal time dependence implicit)
\be\label{comm}
\{A_\a^i({\bf x}),\,E_j^\b({\bf y})\}=\rmi G\delta_\a^\b\delta_j^i\delta({\bf x}- {\bf y})\,.
\ee
Introducing the ``magnetic'' field and the gauge field strength
\ba
B^{\a i}&\equiv& \tfrac12\e^{\a\b\g}F_{\b\g}^i\,,\\
F_{\a\b}^k&=& \p_\a A_\b^k-\p_\b A_\a^k+\e_{ij}\phantom{}\phantom{}^{k}A_\a^iA_\b^j\,,\label{F}
\ea
one can show that the Hamiltonian scalar constraint following from Eq.~\Eq{act} is
\ba
{\cal H}&\equiv& \e_{ijk} E^i\cdot E^j\times \left(B^k+\frac{\Lambda}3E^k\right)\nonumber\\
&=& \e_{ijk} E^{\a i}E^{\b j} \left(F_{\a\b}^k+\frac{\Lambda}{3}\e_{\a\b\g}E^{\g k}\right)=0,\label{sca}
\ea
where $\Lambda$ is a cosmological constant (of any sign) and $\times$ is the vector spatial product defined as $({\bf a}\times {\bf b})_\a= \e_{\a\b\g}a^\b b^\g$. In exterior algebra notation, the gauge field is $A\equiv A_\a \rmd x^\a\equiv A_\a^i \tau_i \rmd x^\a$ ($\tau_i$ being an $su(2)$ generator), the covariant derivative is $D=\rmd+A\wedge$, and Eq.~\Eq{F} can be compactly recast as $F=\rmd A+A\wedge A$. Under a local gauge transformation the Ashtekar connection transforms as
\be\label{gt}
A \rightarrow A'= gAg^{-1} -g^{-1}\rmd g\,,
\ee
where $g({\bf x})$ is an element of the gauge group of gravity $\mathcal{G}=SU(2)$. Let $\mathcal{G}_0$ be the subgroup of \emph{small gauge transformations}, i.e., local transformations continuously connected to the identity. Its elements are of the form $g_0=\exp[-\rmi\tau_i\theta^i(\mathbf{x})]$, where $\theta^i(\mathbf{x})$ are some functions on a spatial slice of ${\cal M}_4$. Pure gauge configurations $g^{-1}\rmd g$ are equivalent to the flat gauge $A=0$.

The full invariance group of the theory is the semidirect product of the diffeomorphism and gauge groups. Invariance under small gauge transformations is guaranteed by the Gauss constraint:
\be
{\cal G}_i
\equiv D_\a E^\a_i=\p_\a E^\a_i+ \e_{ijk}A_\a^j E^{\a k}=0\label{gau},
\ee
while spatial diffeomorphism invariance is imposed by the vector constraint
\be
{\cal V}_\a \equiv (E_i\times B^i)_\a =  F_{\a\b}^iE_i^\b=0\,\label{vec}.
\ee
The total Hamiltonian is a linear combination of the constraints: up to constants, $H=G^{-1}\int_{{\cal M}_3} \rmd^3x (N{\cal H}+\lambda^j{\cal G}_j+N^\a{\cal V}_\a)$, where ${\cal M}_3$ is the spatial submanifold and $N$, $\lambda^j$, and $N^\a$ are Lagrange multipliers (in particular, $\lambda^j$ is a generator of $su(2)$). It can be shown that a classical solution of all constraints is de Sitter spacetime \cite{smo02}, with
\be
A_\a^{i}=\rmi\sqrt{\frac{\Lambda}3}\,\rme^{\sqrt{\Lambda/3}\, t}\delta_\a^i\,,\qquad E_i^\a=\rme^{2\sqrt{\Lambda/3}\, t}\delta^\a_i\,.
\ee
In this family of solutions, parametrized by $\Lambda$, the spin connection $\Gamma_\a^i[E]$ vanishes and the electric and magnetic fields are proportional to each other. Finally, the equation of motion for the connection is
\ba
\dot A_\a^i &=& \{A_\a^i,\,H\}\nonumber\\
            &=& \rmi\e^i_{\phantom{i}jk}E^{\b j}\left(F_{\a\b}^k+ \frac{\Lambda}2\,\e_{\a\b\g} E^{\g k}\right)\,,\label{eom}
\ea
where we have chosen the gauge
\be\label{gei}
N=1/2,\qquad \lambda^i=0,\qquad N^\a=0\,.
\ee
To the action \Eq{act} one can add topological terms which do not affect classical dynamics:
\be\label{actt}
S_{\tilde\t}=\frac{\tilde\t}{8\pi^2}\int_{{\cal M}_4} (*R^{IJ}\wedge R_{IJ}+\rmi R^{IJ}\wedge R_{IJ})\,,
\ee
where $\tilde\t$ is an arbitrary parameter, the first term is Euler class and the second is Pontryagin class. While the Gauss and vector constraints are unchanged,\footnote{The extra term $B_i\times B^i$ in Eq.~\Eq{vec} vanishes identically.} the Hamiltonian constraint \Eq{sca} is modified accordingly \cite{mon01} and the canonical momentum is shifted by a term proportional to the magnetic field:
\be\label{shift}
E^i_a\rightarrow E^i_a-\frac{\rmi\tilde\t}{2\pi^2} B^i_a.
\ee
Then, 
\be\label{shift2}
{\cal H}[E,A]\to {\cal H}_\t= {\cal H}\left[E-\frac{\rmi\tilde\theta}{2\pi^2} B,A\right]\,.
\ee
As $B$ is an axial vector, there appear parity-violating terms at linear and quadratic order in $\tilde\t$.


\section{The Chern--Simons state}\label{csstate}

We move to the quantum level and pick a particular state which solves all quantum constraints. In the quantization of loop quantum gravity in connection representation, the triad and connection operators define the elementary commutation relations \Eq{comm} with $\rmi\hbar\{\cdot,\cdot\}\to[\cdot,\cdot]$, so that
\be
E^\a_i\to -\ell_{\rm Pl}^2\frac{\delta}{\delta A_\a^i}\,,
\ee
while $\hat A_\a^i$ acts as a multiplicative operator; here $\ell_{\rm Pl}^2=\hbar G$ is the Planck area. The reality conditions amount to selecting only the real-valued part of the spectrum of the non-hermitean operator $\hat E$. The quantum constraints on a kinematical state $\Psi(A)$ read
\ba
\hat{\cal H}\Psi(A) &=& \lp^4 \e_{ijl}\e_{\a\b\g} \frac{\delta}{\delta A_{\a i}}\frac{\delta}{\delta A_{\b j}} \hat{\cal S}^{\g l}\Psi(A)=0,\label{qh}\\
\hat {\cal G}_i\Psi(A) &=& -\lp^2D_\a \frac{\delta}{\delta A_{\a i}}\Psi(A)=0,\label{qg}\\
\hat{\cal V}_\a\Psi(A)&=& -\lp^2\frac{\delta}{\delta A_\b^i}F_{\a\b}^i\Psi(A)=0,\label{qv}
\ea
where
\be\label{esse}
\hat{\cal S}^{\g l}\equiv \hat B^{\g l}-\frac{2\pi}{k}\frac{\delta}{\delta A_{\g l}}\,,
\ee
the level $k$ is
\be\label{k}
k\equiv\frac{6\pi}{\ell_{\rm Pl}^2\Lambda}\,,
\ee
and we adopted a factor ordering with the triads on the left \cite{ash86,BGP3}, which makes the quantum constraint algebra consistent.\footnote{This means that $[\hat C_i,\hat C_j]=\hat{\cal O}_{ij}^k \hat C_k$, where $\hat C_i$ are the Hamiltonian, Gauss, and vector constraints and $\hat {\cal O}$ is an operator defining the structure coefficients of the algebra.} A candidate solution to all constraints is the Chern--Simons state \cite{smo02,jac85,kod88,kod90}\footnote{See in particular Exercise 3.7 in \cite{jac85}, p.~258.}
\be\label{cs}
\Psi_{\rm CS}={\cal N} \exp\left(\frac{k}{4\pi}\int_{{\cal M}_3} Y_{\rm CS}\right)
\ee
in Lorentzian spacetime ($k\to \rmi k$ for Euclidean signature, $k\in \mathbb{R}$), where $Y_{\rm CS}$ is the Chern--Simons form \cite{CS}
\ba
Y_{\rm CS}&\equiv& \frac12{\rm tr}\left(A \wedge \rmd A+\frac23 A \wedge A \wedge A\right)\nonumber\\
&=& \rmd^3x\,\e^{\a\b\g}\left(A_\a^i \p_\b A_{\g i}+\tfrac13 \e_{ijk}A_\a^i A_\b^j A_\g^k\right)\,.\label{ycs}
\ea
The trace is taken in the adjoint representation, in which the $su(2)$ algebra generators are $(\e_i)_{jk}=\e_{ijk}$.

The normalization factor ${\cal N}$ is independent of $E$ and $A$, in the sense that it is invariant under infinitesimal transformations of the canonical variables; in general it can depend on the topology, as will be soon demanded. Invariance of the exponent in Eq.~\Eq{cs} under small gauge transformations requires the integration manifold ${\cal M}_3$ to have no boundary. As the classical spatial topology is determined by a superposition of quantum states, one should sum over inequivalent closed three-dimensional Cauchy surfaces compatible with gauge invariance. In \cite{soo01} it is argued that ${\cal M}_3$ must have constant curvature in the semi-classical limit, selecting the 3-sphere $S^3$ as the only classically viable alternative.

Solution of the Hamiltonian constraint is granted by the operatorial ordering of Eq.~\Eq{qh} and the property
\be
\frac{\delta}{\delta A_{\g k}}\int_{{\cal M}_3} Y_{\rm CS}=2B^{\g k}\,.
\ee

The vector constraint as written in Eq.~\Eq{qv} is not solved by $\Psi_{\rm CS}$, which is in the kernel of $\hat F\hat E$. One can regularize it with an even smearing function $f_\varepsilon({\bf x}-{\bf y})$ on ${\cal M}_3$ such that $\lim_{\varepsilon\to 0} f_\varepsilon({\bf x}-{\bf y})=\delta({\bf x}-{\bf y})$ (e.g., \cite{GP}; $\lp$ factors omitted):
\be
\int \rmd^3xN^\a{\cal V}_\a=\lim_{\varepsilon\to 0} \int \rmd^3x\rmd^3y N^\a({\bf x}) f_\varepsilon({\bf x}-{\bf y})\frac{\delta}{\delta A_\b^i({\bf x})}F_{\a\b}^i({\bf y}).
\ee
The functional derivative of $F$ does not give any contribution, as 
\ba
\int \rmd^3 y f_\varepsilon\frac{\delta F_{\a\b}^i}{\delta A_\b^i}&=&\int \rmd^3 y \p_\b [\delta({\bf x}-{\bf y}) f_\varepsilon({\bf x}-{\bf y})]\nonumber\\
&&+\int \rmd^3 y f_\varepsilon \e_{jk}\phantom{}\phantom{}^{i} (\delta_\a^\b\delta_i^jA_\b^k+\delta_i^kA_\a^j)
=0\nonumber
\ea
by parity of $f_\varepsilon$.

Given the semblance of the Ashtekar formulation to Yang--Mills theory, it is tempting to seek a relation between the Chern--Simons form and a fermionic current $J$ similar to that evaluated in the triangle diagrams in Yang--Mills theory coupled to fermions, $\rmd*J\sim Y_{\rm CS}$.\footnote{In a similar way, the Immirzi parameter emerges from the topological sector of gravity \cite{GOP,mer07}.} In what follows we shall find an analogous situation but inclusion of fermions by hand will be replaced by a deformation of the $\t$ sector. Fermions will emerge from the spin-network embedding of the Chern--Simons state.

The full gauge group of gravity $\mathcal{G}$ contains also \emph{large gauge transformations} \cite{CDG,JR}, namely, unitary transformations $g(\mathbf{x})$ which are not homotopic to the identity but tend to the identity at large $\mathbf{x}$, $\lim_{|\mathbf{x}|\to\infty}g(\mathbf{x})=\mathbb{I}$. All directions in Euclidean space are identified at infinity, so $g$ is a mapping from ${\cal M}_3\cong S^3$ to $S^3$. The different ways in which the sphere $S^3$ can be continuously mapped onto itself are summarized by the homotopy group $\pi_3(S^3)=\mathbb{Z}$. 
Therefore each disconnected component of ${\cal G}$ is labelled by an integer $n$, called \emph{winding number}, generated out of a pure gauge connection $A^o=g^{-1}\rmd g$ in that component,
\be 
w(A^o)=\frac{I(A^o)}{24\pi^2}=\frac{1}{24\pi^2}\int_{S^3} {\rm tr}(A^o\wedge A^o \wedge A^o),
\ee
where $I$ is the Cartan--Maurer invariant \cite{wei}. The winding number characterizes how many times $g(\mathbf{x})$ winds around the noncontractible 3-sphere in the $SU(2)$ internal space as $\mathbf x$ ranges over $S^3$ in space. If $g=g_0\in{\cal G}_0$ the winding number is zero, while all large gauge transformations fall into homotopy classes $g_n$ generated by the $n=1$ transformation \cite{JR}: $g_n(\mathbf{x})=[g_1(\mathbf{x})]^n$. Then the quotient $\mathcal{G}/\mathcal{G}_0$ is isomorphic to the homotopy sequence of integer-valued winding configurations $\{(g_1)^n\,|\,n \in\mathbb{Z}\}$.

This is the origin of topological $\theta$ vacua in Yang--Mills theory \cite{CDG,JR} and gravity \cite{DDI,ABJ}. On each disconnected component of ${\cal G}$ there lives an independent physical sector (an inequivalent quantum theory) with ground state $\Psi(A)=\langle A|n\rangle$. A unitary large gauge transformation jumps from a sector to another, acting as $\hat g_1|n\rangle=|n+1\rangle$. By definition, physical observables are invariant under the full gauge group, so $\hat g_1$ must commute with the Hamiltonian and they share the same basis of eigenstates, chosen so that the eigenvalue of $\hat g_1$ on a state is a pure phase $\rme^{\rmi\t}$, $0\leq \t\leq 2\pi$. In other words, under a gauge transformation as in Eq.~\Eq{gt}, $\Psi(A')=\Psi(A)$ if $g=g_0$ is small, while $\Psi(A')=\rme^{\rmi n\t}\Psi(A)$ if $g=g_n$ is large. This implies that the quantum vacuum is a superposition of states $|\t\rangle =\sum_n \rme^{\rmi n\t}|n\rangle$.

The Chern--Simons form transforms nontrivially under large gauge transformations, as
$\int Y_{\rm CS}\to \int Y_{\rm CS}+4\pi^2w$:
\be\label{u1}
\Psi_{\rm CS}(A)\to \Psi_{\rm CS}(A')=\rme^{\rmi n\t}\Psi_{\rm CS}(A)\,,\qquad n\in\mathbb{Z}\,,
\ee
where
\be\label{thk}
\rmi\t= k\pi=\frac{6\pi^2}{\ell_{\rm Pl}^2\Lambda}\,.
\ee

The $\t$ parameter is indeed a phase in Euclidean case, where istantonic solutions are considered. Invariance of the state under large gauge transformations can be achieved by taking $\Psi'(A)\equiv  \rme^{-\frac{\rmi\t}{4\pi^2}\int Y_{\rm CS}(A)}\Psi_{\rm CS}(A)={\cal N}$, a constant functional; then the momentum $\hat E$ is transformed into \cite{ABJ} $\hat E'=\Psi_{\rm CS}^{-1}\hat E\Psi_{\rm CS}$, which is nothing but Eq.~\Eq{shift} with $\tilde\t=\lp^2\t$. The $\theta$ dependence is transferred to the Hamiltonian, which now acts on $\Psi'(A)={\rm const.}$ as a connection derivative and is trivially annihilated: $\hat{\cal H}_\t\Psi'\propto \hat E'\hat E' \delta{\cal N}/\delta A=0$. This simple result is due to the fact that the state under consideration is also the unitary rotation transforming the triad operator. Another way to achieve state invariance is by setting the normalization constant to be \cite{soo01,PG}
\be
{\cal N}=\rme^{-\rmi w(A^o)\t},
\ee
which does not require modifications of the constraints (as is the case for general states \cite{ABJ}). Note that the $\theta$ dependence is absorbed in the state normalization but reappears in the inner product, giving inequivalent quantum probabilities.


\subsection{Open issues}

The Chern--Simons state has been proposed to define a natural clock for gravitational systems \cite{SS}. Moreover, it is a genuine ground state of the theory, inasmuch as by adding a matter sector its weakly-coupled excitations reproduce standard quantum field theory on de Sitter background \cite{SS}, while linearizing the quantum theory one recovers long-wavelength gravitons on de Sitter \cite{smo02}. However, at this point we should mention several issues raised against the viability of this state. 
\begin{itemize}
\item The first is that $\hat{\cal H}$ as given in Eq.~\Eq{qh} is not an hermitean operator and, on the other hand, $\Psi_{\rm CS}$ is not a solution of the constraint for other operator orderings, like that with triads on the right \cite{JS}.
\item As the connection $A$ cannot be promoted to a well-defined operator, one would like to reexpress the Chern--Simons state in terms of holonomies $h[\Gamma,A]$, where $\Gamma$ is a set of loops upon which integration of the holonomy is carried out. In the loop representation of loop quantum gravity \cite{GT,RS1,RS2} (see \cite{smo02} for a review), $\Psi_{\rm CS}$ can be written as $\Psi(\Gamma)=\langle h[\Gamma,A],\Psi(A)\rangle$, that is, as an integral in connection space with a suitable measure $\mu(A)$ \cite{BGP1,BGP2}. When $A$ is real, $\Psi(\Gamma)$ was shown \cite{wit89} to be a topological invariant, a knot polynomial in the variable $\Lambda$ called \emph{Kauffman bracket} \cite{smo02,GP,kau91,KL,BM}. This identification would open up an intriguing possibility of describing quantum gravitational dynamics with the mathematics developed in knot theory \cite{GP,kau91,KL,KLo2}. Notwithstanding, there are a number of missing rigorous steps in the beautiful picture sketched so far. First, a definition of the measure $\mu(A)$ in configuration space is still under construction, and the scalar product $\Psi(\Gamma)$ is only formal.\footnote{An inner product in configuration (connection) space was suggested in \cite{SS}.} Second, knot theory is based upon a real gauge variable, while the Chern--Simons state \Eq{cs} is constructed with the self-dual connection.
\item $\Psi_{\rm CS}$ is not normalizable. This is expected on the ground that states satisfying the constraints are generally non-normalizable in the inner product on the kinematical Hilbert space; also, some connection components act as time coordinates on the configuration space, while the physical inner product does not integrate over time (lest energy eigenstates be non-normalizable) \cite{smo02,SS}.
\item $\Psi_{\rm CS}$ violates CPT \cite{wit03}, which may result in negative energy levels and Lorentz violation. 
\end{itemize}
To the best of our knowledge, one can address these points as follows. In principle one might absorb the triad prefactor operator $\hat E \hat E$ in a new measure in configuration space but we do not need to resort to this trick, as hermiticity is a requirement only for excited states in a quantum mechanical theory. Since in gravity with no matter modes one is interested in the ground state only, the first problem is presently immaterial. Also, we will start from a degenerate sector where $\hat {\cal H}=0$ identically, so this constraint will play no dynamical role. Rather, we will be interested in the ``reduced'' constraint Eq.~\Eq{esse}.
 
Despite the incompleteness of its formulation, the loop transform is a meaningful manipulative tool to obtain results that can be verified, in several independent ways, by nontrivial perturbative consistency checks, an account of which is given in \cite{GP} (Sect.~10.5.2-3). A recent example is given by \cite{soo01}, where the formal analogy between the Chern--Simons state and the Kauffman bracket has been made somewhat more concrete. To switch from connection to triad representation \cite{ABJ}, one makes use of an inversion formula defined via an integration contour of the connection along the imaginary axis \cite{kod90}. A consequence is that the connection variable can be treated as real after a Wick rotation, thus linking to the results of knot theory: in fact, one can resort to perturbative techniques almost identical to the real-connection case. 

Further, the Chern--Simons state has been generalized to a real Immirzi parameter (real connection) in \cite{ran1,ran2}. This state solves all the quantum constraints, is $\delta$-normalizable (since de Sitter space is unbounded, the wavefunction corresponding to it cannot be normalizable \cite{ran3}), invariant under large-gauge and CPT transformations (it violates CP and T separately), and coincides with the Kauffman bracket in loop representation.

Although there is no rigorously known (kinematical) Hilbert space for self-dual gravity with Lorentzian signature, one can still make generic predictions about the quantum theory. In fact, many key features of quantum mechanics and quantum field theory were discovered before a Fock or Hilbert space were defined. The example of momentum eigenstates in quantum mechanics, which are only $\delta$-normalizable, well illustrates the case of the Chern--Simons state \cite{ran3}.

The Chern--Simons state can give a useful insight in nonperturbative properties of the theory. In other words, we believe that the scenario we shall propose is suggestive of a framework that is \emph{independent} of the problems associated with the state. As for the moment we are not in a position to prove our belief, we wish to make a last remark. One of the properties of the Chern--Simons state is that it is both an exact state and a WKB state. Thus, even if one does not believe the state to be in the Hilbert space of the theory, or can be rigorously defined in the framework of LQG, one can still argue that at some level a true quantum state that is close to the de Sitter ground state must be reasonably well approximated by the Chern--Simons state (up to some order in $\hbar$). So one cannot simply dismiss the state even if one does not believe it really exists in the physical Hilbert space.

For all these reasons, we are encouraged to make use of the Chern--Simons state in the present investigation, adding our findings to the positive evidence, still an open subject of discussion, in favour of its relation with some yet unexplored sector of quantum gravity.


\section{Deformed topological sector and Chern--Simons state}\label{defo}

Whenever a cosmological constant component is required by observations at some point during the evolution of the universe, several fine-tuning issues lead to the phenomenological hypothesis that the vacuum component is actually dynamical; its behaviour can be reproduced by the matter (typically, scalar) field characteristic of quintessence and inflationary models.

In this respect what we are going to do, promoting $\Lambda$ to an evolving functional $\Lambda(A)$, crosses a well-known path. Nevertheless, the justification and consequences of this step change perspective under the lens of loop quantum gravity.

First, the operation 
\be
\Lambda\to\Lambda(A)
\ee
is recognizable, thanks to Eq.~\Eq{thk}, as a deformation of the topological sector of the quantum theory:
\be
\t\to\t(A)\,.
\ee
From the Lagrangian perspective, the appearence of a $\t$ sector in the Hamiltonian is tantamount (on solutions to the field equations) to adding the self-dual Pontryagin term Eq.~\Eq{actt}. The same argument holds for Yang--Mills theory, but because of the Adler--Bell--Jackiw anomaly \cite{A,BJ} we can freely rotate this effect away by a chiral phase redefinition of massless quarks \cite{tH1,tH2}. In QCD the partition function can be extended to a larger $U(1)$ symmetry, namely the Peccei--Quinn symmetry corresponding to invariance under a rotation by the $\theta$ angle \cite{PQ1,PQ2}. Instanton effects can spontaneously break this $U(1)$ symmetry resulting in a light particle called axion. 

Likewise, a large gauge transformation in Chern--Simons wavefunction is regarded as a $U(1)$ rotation, Eq.~\Eq{u1}, so by deforming $\theta$ we allow $\theta$ to vary as a function of the connection, which explicitly breaks the $U(1)$ symmetry.\footnote{To deform $\Lambda$ by scalar matter was early recognized as breaking conformal invariance of de Sitter spacetime \cite{CoV}.} The gravitational Hamiltonian is augmented by a conterterm breaking gauge invariance and the Kodama state gets a phase modification that is dynamically equivalent to a fermion bilinear operator. The flux configuration in the $\theta$ vacua is equivalent to Jacobson's gauge configuration in $1+1$ dimensions, establishing a sort of holographic duality taking place on the horizons which connect spacetimes in different winding sectors.

Since the cosmological constant is deformed in the Hamiltonian constraint, the corresponding Chern--Simons state, $\Psi_*$, will also have to be deformed because, according to Eqs.~\Eq{u1} and \Eq{thk}, $\Lambda$ enters in the phase of the wavefunction. As the Chern--Simons state lives in the connection space, we assume that $\Lambda$ in the deformed state $\Psi_*$ depends only on $A$ and not on the triad operator, and lies inside the integral. 

The classical constraint algebra determines the functional dependence of $\Lambda$. Denoting with $H_*=\int \rmd^3xN{\cal H}_*$ the Hamiltonian constraint with deformed $\Lambda$, to get gauge-invariant dynamics we demand that its Poisson bracket with the Gauss constraint ${\cal G}=\int \rmd^3x \lambda^i {\cal G}_i$ vanishes:
\be
\{H_*,\mathcal{G}\}=0\,,
\ee
yielding
\be
\{\Lambda(A),\,{\cal G}\}\det E =0\,.
\ee
The Poisson bracket between $\Lambda$ and ${\cal G}$ does not vanish unless $\Lambda$ is invariant under small gauge transformations, as
\be
\{\Lambda(A),\,{\cal G}\}=\frac{\delta \Lambda}{\delta A_\a^i}\frac{\delta {\cal G}}{\delta E^\a_i}=0 \quad \Leftrightarrow\quad
  -\frac{\delta \Lambda}{\delta A_\a^i} D_\a\lambda^i=0\,,\label{x}
\ee
where we have integrated by parts the Gauss constraint. Another possibility, which does not exclude the former, is $\det E=0$, leading to a degenerate sector of gravity with ${\rm rk} E\leq 2$ (rk denotes the rank of the triad). We ignore the most degenerate case ${\rm rk} E=0$, as we want to preserve at least part of the canonical algebra. This leaves the cases ${\rm rk} E=1,2$.

The quantum scalar constraint $\hat{\cal H}_*$ is defined with $\Lambda(A)$ to the left of triad operators. One can assume the same attitude as in the discussion of the Chern--Simons state, and require that $\Psi_*$ annihilates the deformed reduced constraint ${\cal S}_*^{\a i}$, Eq.~\Eq{esse}. The latter requires the addition of a counterterm $\hat \Theta^{\a i}$:
\be
(\hat\Theta^{\a i}+\hat{\cal S}_*^{\a i})\Psi_*=\hat \Theta^{\a i}+\frac12\int_{{\cal M}_3} Y_{\rm CS}\frac{\delta \ln \Lambda(A)}{\delta A_{\a i}}=0\,,\label{th0}
\ee
leading to
\be\label{th}
\hat \Theta^{\a i}=\frac1{2\lp^2}\int_{{\cal M}_3} Y_{\rm CS} \hat E^{\a i} \ln \Lambda.
\ee
The quantum, nonperturbative origin of this term is clear by the presence of the $1/\lp^2$ prefactor. Equation \Eq{th} breaks invariance under large gauge transformations, since the deformation of $\Lambda$ is actually a change of the topological sector. Taking Eq.~\Eq{th0} at face value as a classical expression, Eq.~\Eq{eom} is modified as
\be\label{eom2}
\dot A_\a^i = \rmi\e^i_{\phantom{i}jk}E^{\b j}\left[F_{\a\b}^k+ \frac{\Lambda}2\,\e_{\a\b\g} E^{\g k}
-\e_{\a\b\g}\int_{{\cal M}_3} Y_{\rm CS}\frac{\delta \ln \Lambda(A)}{\delta A_{\g k}}\right]\,.
\ee
We have noticed that a dynamical cosmological constant consistent with the vacuum quantum constraints can be accommodated in a degenerate sector of gravity. We shall develop this picture and its consequences in the special case ${\rm rk} E=1$. This is classical Jacobson's configuration \cite{jac96}, reviewed in the next subsection.

If the triad operator is degenerate the scalar constraint is trivial and nondynamical. In particular, there are no anomalies in the degenerate constraint algebra \cite{BoKS}. The choice of Jacobson's sector only on account of this reason may seem too drastic without further inspection of the algebra. Actually, not only the $(1+1)$-dimensional model is under good control from a technical and physical point of view, but also it will allow us to give simple ``holographic'' reinterpretations of, for instance, the cosmological constant, geometric measurements, and the boundary entropy of a de Sitter spacetime. This is our main motivation for considering ${\rm rk} E=1$. The case ${\rm rk} E=2$ will be dealt with elsewhere.


\subsection{Jacobson classical degenerate sector}\label{jaco}

One of the advantages of the formulation of classical gravity via Ashtekar variables is the possibility to have a well-defined causal structure even when the background metric is singular, as inside a black hole, at big-bang or big-crunch events, or in processes where topology changes \cite{ho91a,ho91b}. Degenerate sectors with causal structure were classified, at the classical level, in \cite{mat95,LW1,LW2}. For every point $x\in{\cal M}_4$, one defines the \emph{future} as the set of all tangent vectors at $x$ which are images of positive timelike or lightlike vectors in $M_4$:
\be
{\cal F}(x)=\left\{\zeta^\mu(x)=\zeta^IE_I^\mu(x)\in{\cal M}_4\,\,\big|\,\,\zeta_I\zeta^I\leq 0,\,\zeta^0>0\right\}\,,
\ee
where $\zeta^I\in M_4$. The \emph{past} is defined as $-{\cal F}(x)$. Causality is a meaningful concept if the future and the past of a given point in ${\cal M}$ do not intersect \cite{mat95},
\be\label{cau}
-{\cal F} \cap{\cal F}=\emptyset\qquad \Leftrightarrow\qquad 0 \notin {\cal F}\,,
\ee
which guarantees the existence of a spacelike surface at any point in spacetime (a surface is spacelike if it does not intersect with ${\cal F}$, i.e., if it only contains causally disconnected points).

Enforcement of the causal condition \Eq{cau} determines the future submanifold according to the rank of the tetrad. For an invertible metric (${\rm rk} E=4$, rk being the rank of the tetrad), ${\cal F}$ is a hypercone without tip, while for ${\rm rk} E=3$ it is a tipless cone. When ${\rm rk} E=2$, the future is a tipless wedge. Finally, for ${\rm rk} E=1$ or 0 ${\cal F}$ is a half line or $\{0\}$, respectively.

Here we are interested in the rank-2 case and, in particular, in the classical field configuration described by Jacobson \cite{jac96} (sector $(1,1)$ according to the nomenclature established in \cite{LW2}). Geometry amounts to a collection of gravitational lines where the triad $E^\a_i$ has rank 1 and vanishes elsewhere. On a line embedded in a local coordinate system $\{x\}$,
\be\label{deg}
E^\a_i(x)=V^\a(x)\tau_i(x)\,,
\ee
where $V^\a$ is the vector tangent to the line at $x$ and $\tau_i$ is a vector in the internal gauge space. The scalar constraint \Eq{sca} is identically satisfied thanks to Eq.~\Eq{deg}.

The reality conditions Eq.~\Eq{rc} imply that both $E_i^\a E^{i\b}$ and $\dot E_i^\b$ are real \cite{ART}, thus having a straightforward extension to the degenerate sector. In particular, one can choose a real vector $V^\a\in\mathbb{R}^3$ and $\tau_i\tau^i>0$. Spatial diffeomorphism invariance can be partially fixed by taking the vector $\mathbf{x}=(x^1,x^2,z)=(x^a,z)$ and spreading the gravitational line along the $z$ direction. Then $V^\a=(0,0,1)^\a$ and the only nonzero component of the electric field is $E^z_i$. The Gauss constraint Eq.~\Eq{gau} becomes $D_z\tau^i=0$, stating that $\tau^i$ is parallel transported along the line. Therefore one can take a gauge where the $z$ component of the connection is constant on the curve:
\be
A_z^i= f(x^a)\tau^i(t,x^a,z),
\ee
where $f$ is a function of the transverse coordinates. After further fixing the diffeomorphism gauge to $\tau^i=(0,0,\tau)^i$, the vector constraint \Eq{vec} reads
\be
\p_zA_a^3-\p_a (f\tau)=0\,.
\ee
If the curve is closed ($z$ periodic) the only possible classical solution is $f\tau={\rm const}$, and the holonomy is the same in all $z$ loops for a given $x^a$. Barring this configuration, one is entitled to choose the gauge $f=0$ and Eq.~\Eq{gei}, leading to
\bs\label{des}\ba
&& A_z^i=0\,,\qquad A_a^3=A_a^3(x^a)\,,\qquad A_a^{i\neq 3}=A_a^{i\neq 3}(t,z)\,,\\
&& E^z_{i\neq 3}=0\,,\qquad E^a_i=0\,,\qquad E^z_3=E^z_3(x^a)\,,
\ea\es
where the last equality stems from the Gauss constraint and the equations of motion (the presence of a cosmological constant in the Hamiltonian constraint is irrelevant). The only vanishing components of the magnetic field are $B_a^3$, $a=1,2$.

The residual freedom in the choice of the transverse coordinates $x^a$ can be used to fix $\tau=1$ ($E^z_3=1$). The equation of motion \Eq{eom} for the transverse-transverse components of the connection is $\dot A_a^i= -\rmi\epsilon^{i3}_{\phantom{i3}j} \p_z A_a^j$, which can be written as two Weyl equations
\be\label{45}
\dot\psi_\pm+\s^z\p_z\psi_\mp=0\,,
\ee
where
\be
\psi_+\equiv\bma{c} \rmi A^1_1 \\ A^1_2\ema\,,\qquad \psi_-\equiv \bma{c} A^2_1 \\ \rmi A^2_2\ema\,,
\ee
and $\sigma^z$ is the third Pauli matrix:
\be
\sigma^x =\bma{cc}0\,&1\\1\,&0\ema\,\quad \sigma^y=\bma{cc} 0&-\rmi\\\rmi&0\ema,\quad  \sigma^z =\bma{cc}1&0\\0&-1\ema.
\ee
Other field definitions are possible; for instance, one can consider the two two-spinors $(A^1\pm \rmi A^2)_a$, which propagate at light speed in opposite directions along the gravitational line \cite{jac96}. Here, we rather take the four-spinor
\be\label{psia}
\psi\equiv\bma{c} \psi_+ \\ \psi_-\ema\,,
\ee
for which one can write, at every point $x^a$ in the trasverse plane, the $(1+1)$-dimensional Dirac equation
\be\label{dieq}
\g^0\dot\psi+\g^z\p_z\psi=0\,,
\ee
where
\be
\g^0 =\bma{cc}\mathbb{I}_2&0\\0&-\mathbb{I}_2\ema\,,\qquad \g^\a=\bma{cc}0&\s^\a\\-\s^\a&0\ema
\ee
are Dirac matrices. The causal structure of the degenerate sector, Eq.~\Eq{dieq}, is that of a ``worldsheet'', a $(1+1)$-dimensional spacetime. Rotating back in the target space, Eq.~\Eq{dieq} can be put in covariant form as $\g^\mu\p_\mu\psi=0$.

A Yang--Mills field gives no extra contribution to the analysis; a Klein--Gordon scalar field would be seen as a constant of motion along the gravitational lines, while a matter fermion would propagate on the worldsheet only.

One can take several differently oriented gravitational lines and patch them together at their boundaries \cite{jac96}. The emerging classical picture is that of two-dimensional worldsheets interacting at the edges, where the gravitational field may be nondegenerate. However, one lacks a model for this interaction, as well as a physical interpretation of the worldsheet network and fermionic degrees of freedom. 

Both naturally emerge at the quantum level when the operatorial constraints of the previous section act upon the Chern--Simons state. As is expected, one can identify a conformal field theory living on the worldsheets, while conformal invariance is broken at the boundary by a marginal operator encoding the interaction. As the classical picture suggests, this deformed CFT is fermionic in nature and, more precisely, describes a Fermi liquid. On the other hand, its geometric and algebraic structures resemble those of a quantum (or framed) spin network which is, not surprisingly, the natural environment wherein to embed the Chern--Simons state. This piece of evidence leads to the conjecture that the degenerate sector has a direct interpretation in terms of framed spin networks, so that the $(1+1)$-dimensional theory holographically generates the \emph{full} four-dimensional spacetime.


\subsection{Quantum gravity as a Fermi-liquid theory}\label{main}

Jacobson equivalence is what is needed to establish the emergence of massive fermions from winding transitions in the Chern--Simons state. Normally, degenerate metrics signal a singularity (as in cosmological spacetimes) and horizons. Degenerate spaces also allow for topology transitions. We will show that a solution of the deformed equation of motion has a cosmological constant of the form
\be\label{mainl1}
\Lambda=\Lambda_0 \exp(-j^{5z})^n\,,
\ee
where $\Lambda_0$ is a natural integration constant of order unity, $n>0$ is a constant,
\be\label{mainl2}
j^{5z}\equiv\bar\psi\g^5\g^z\psi
\ee
is a fermionic axial current and
\be
\g^5\equiv \rmi\g^0\g^x\g^y\g^z=\bma{cc}0&\mathbb{I}_2\\\mathbb{I}_2&0\ema\,.
\ee

Let us go back to the effective Hamiltonian of quantum gravity with the nonperturbative counterterm. In Jacobson's sector, the transverse-transverse connection components behave as fermionic degrees of freedom and Eq.~\Eq{eom2} becomes
\be\label{eom3}
\dot A_a^i = -\rmi\e^i_{\phantom{i}3k}\left[\p_z A_a^k+\e_{azb}\int_{{\cal M}_3} Y_{\rm CS}\frac{\delta \ln \Lambda(A)}{\delta A_{b k}}\right]\,,
\ee
where $a,b\neq z$ and $i,k\neq 3$. The extra term can be manipulated to express it as a function of $\psi$ fields. In fact, the Chern--Simons functional is
\ba
Y_{\rm CS}&=&-(\psi_+^T\s^y\p_z\psi_+ + \psi_-^T\s^y\p_z\psi_-)\,\rmd^3x\nonumber\\
&=&\rmi(\psi^T\g^x\g^z\p_z\psi)\,\rmd^3x\,,\label{ycspsi}
\ea
where $T$ denotes the transpose. To proceed, one has to make an ansatz for the functional $\Lambda(A)$. We choose
\be
\Lambda(\psi)=\Lambda_0 \exp (\xi^{T}\psi)^n\,,
\ee
where $\xi^{T}=(\xi_1,\xi_2,\xi_3,\xi_4)$ is, for the time being, arbitrary. Then
\be
\e_{i3k}\e_{azb}\frac{\delta \ln \Lambda[A({\bf x})]}{\delta A_{b k}({\bf y})}=n(\xi^{T}\psi)^{n-1} \Xi_{ia}\delta({\bf x}-{\bf y})\,,
\ee
where we defined the matrix
\be
\Xi=\bma{cc} \rmi\xi_4 & -\xi_3\\ -\xi_2 & \rmi\xi_1\ema\,.
\ee
Equation \Eq{eom3} yields
\be\label{eom4}
\g^0\dot\psi+\g^z\p_z\psi+\rmi \widetilde m\g^x\xi=0\,,
\ee
where
\be\label{mass}
\widetilde m\equiv -\rmi n(\xi^{T}\psi)^{n-1}(\psi^T\g^x\g^z\p_z\psi)\,.
\ee
This effective mass is nonperturbatively generated by the anisotropic cross-interaction of the connection components. Imposing the condition 
\be\label{dir}
\xi=-\g^x\psi,
\ee
the field $\psi$ would obey a Dirac equation with mass $m=2\widetilde m$. Classically the mass would vanish, as $\xi^T\psi=0$. However, after quantizing the field its components become Grassmann variables and the symmetry-breaking mechanism comes into effect. In order to get quantities with well-defined Lorentz structure, $\psi^T$ should actually be related to $\bar\psi\equiv\psi^\dagger\g^0$. This is realized if $\psi=\psi_c\equiv -\rmi\g^y\psi^*$, i.e., if $\psi$ is equal to its charge conjugate (Majorana fermion). Then the connection components must satisfy the conditions
\be\label{real}
A_1^2=A_2^{1*}\,,\qquad A_2^2=A_1^{1*}\,.
\ee
In particular,
\be
\xi^T\psi=\psi^T\g^x\psi=-j^{5z}\,,
\ee
and
\be\label{mass2}
m\equiv -2\rmi \,n(-j^{5z})^{n-1}\,\bar\psi\g^5\p_z\psi\,.
\ee
Accordingly, the cosmological constant encodes the imprint of an axial vector current, associated with a chiral transformation of the fermion $\psi$, 
\be\label{chir}
\psi\to \rme^{\rmi\t\g^5}\psi,
\ee
and not conserved in the presence of a mass term. This is yet another reminder that the deformation process affects the topological sector and the CP symmetry of the theory. The transformation for individual transverse-transverse connection components sends the latter to others with exchanged indices, $1\leftrightarrow 2$ (Eq.~\Eq{real}).

Note that if one enforced the condition
\be\label{maj}
\xi=\g^x\psi_c\,,
\ee
Equation \Eq{eom4} would be a Majorana equation, and
\be 
\xi^T\psi=j^{5z}\,.
\ee
Up to a sign, this choice is equivalent to the former, to which we shall refer from now on. 

Note that Eq.~\Eq{mainl1} is a particular form of a Lorentz-invariant cosmological constant $\Lambda\propto \exp (-E\cdot j^5)^n$; the factor $E^z_i\tau^i$ in front of $j^{5z}$ is 1 in our gauge conventions, declared above Eq.~\Eq{45}. It would be interesting to extend the analysis to a gauge-invariant approach and investigate these properties without gauge fixing. This issue and the properties of the effective mass Eq.~\Eq{mass2}, as well as the compatibility between Eq.~\Eq{real} and the reality conditions,\footnote{For a dS background the reality conditions would require $A$ to be pure imaginary, but these would be modified since classical solution of the deformed model depends on the form of $\Lambda(A)$.} will require further investigation we will not pursue here.

We propose Eqs.~\Eq{mainl1}, \Eq{mainl2} and \Eq{mass2} as the starting point of a physical reinterpretation of quantum gravity. One can canonically quantize $\psi$ and expand it in discrete one-dimensional momentum space,
\be\label{fou}
\psi = \sum_{k,\s} \frac{\rme^{\rmi kz}}{\sqrt{2\cE_k}} \left(c_{k\s}u_{k\s}+c_{-k\s}^\dagger v_{k\s}\right)\,,
\ee
where $\s=\pm$ is the spin, $\cE_k$ is the energy, $c$ and $c^\dagger$ are annihilation and creation operators obeying a fermionic algebra, and $u$ and $v$ are spinorial functions. Due to Eq.~\Eq{mass}, in the Hamiltonian there appear  fermionic nonlocal correlations of the form
\be
(c^\dagger_{k_1\s_1}c_{k_1'\s_1'}\dots c^\dagger_{k_{n-1}\s_{n-1}}c_{k_{n-1}'\s_{n-1}'}) c^\dagger_{k\s}c^\dagger_{k'\s'}c_{k\s}c_{k'\s'}\,.
\ee
Spin models of this form, generically called Fermi-liquid theories, are employed in condensed matter physics to describe fermionic systems with many-body interactions at sufficiently low temperature (e.g., \cite{Vol1,Vol2,Vol03} for introductory reviews).\footnote{The works of Volovik were the first to apply condensed-matter analyses to quantum particle and gravitational physics and they constitute a precursor in this direction. In particular, they exploited formal analogies between superfluid phases of $^3{\rm He}$ and effective models of cosmic strings and monopoles, gravity and high-energy physics \cite{Vol1,Vol2,Vol03,Vol97,Vo981,Vo982,Vol3,Vol4,Vol5,Vol6}, as well as black holes and degenerate metrics \cite{JV1,JV2,Vol99,FiV,Vo031}. Gravity and the cosmological constant problem were regarded as emergent properties of a superfluid \cite{Vol1,Vol2,Vol03,Vol3,Vol4,Vol5,Vol6}, in a way which considerably differs from ours as far as physical interpretation and mathematical tools are concerned.}

In the remainder of this paper we substantiate the emergent physical picture by taking a simple example in this family of models, given by BCS superconductivity \cite{BCS,bog}, to which we devote the next subsection. BCS theory adopts the choice
\be
n=1
\ee
in the above equations, so that the nonlocal cross-interaction in the Hamiltonian is four-fermionic and pair-wise:
\be\label{vkk}
\sum_{\s,k,k'}V_{kk'}c^\dagger_{k\s} c^\dagger_{k'\s'} c_{k\s} c_{k'\s'}\,,
\ee
where $V_{kk'}$ is a function of the momenta and is determined by the mass $m= -2\rmi\bar\psi\g^5\p_z\psi$. The corresponding structure (a $1+1$ fermionic CFT with boundary terms breaking conformal invariance) will turn out to closely reproduce that of LQG quantum spin networks.

We have not shown, nor we will in this work, whether quantization of the four-component object $\psi$ as a fermion, Eq.~\Eq{fou}, is compatible with the bosonic nature of the connection. The issue of spin-statistics is still wide open, together with the explicit realization of a two-to-four dimensional, BCS-to-canonical gravity transition; see the conclusions. For the moment, we recall an intriguing result in the literature. In \cite{MTR1,MTR2} it was argued that the action of pure-gravity Hamiltonian on ``open loop'' states, where the end points each carry a fermionic degree of freedom, reproduces an Einstein--Weyl theory in the classical limit. This finding suggests a possible way out of the fermion problem. The $1+1$ world of Jacobson's sector might be identified with the one spanned by the endpoints of ``open loops'' in the old loop representation, where fermion fields naturally live. In our case, fermions are not independent from the connection: the column $\psi$ is made of connection components, but with a statistic different from the bulk. Rather than working in loop representation, below we shall begin to attack the problem from the perspective of spin networks.


\subsection{Review of BCS theory}\label{bcs}

Let us pause and describe some general features of reduced BCS superconductivity theory \cite{sie99,AFS,sie01}. An electron with discrete momentum ${\bf k}$, energy $\cE_{\bf k}$, and spin $\s=\pm$ is represented by the state $|{\bf k},\s\rangle=c_{{\bf k}\s}^\dagger|0\rangle$, where $|0\rangle$ is the single-particle vacuum state and $c^\dagger_{{\bf k}\s}$ is the spin-carrying fermion creation operator (obeying the algebra $\{c_{{\bf k}\s},c_{{\bf k}'\s'}^\dagger\}=\delta_{{\bf k}{\bf k}'}\delta_{\s\s'}$). One can consider a Fermi sea of $2N$ electrons and $\Omega\geq N$ doubly degenerate energy levels described by the Hamiltonian (momentum dependence is understood)
\be\label{bsch}
H_{\rm BCS}=\sum_{a,\s} \cE_a c^\dagger_{a\s}c_{a\s}-\sum_{a,a'}V_{aa'}c^\dagger_{a+} c^\dagger_{a-} c_{a'-} c_{a'+}\,,
\ee
where $a=1,\dots \Omega$, the creation and annihilation operators $c^\dagger_{a\s}$, $c^\dagger_{a\s}$ are Fourier transforms of the momentum-space operators and satisfy the anticommuting algebra $\{c_{a\s},c_{a'\s'}^\dagger\}=\delta_{aa'}\delta_{\s\s'}$ (zero structure constants otherwise), and $V_{aa'}>0$ is a nonlocal two-body interaction between the fermionic pairs labelled by $a$ and $a'$, entailing the exchange of virtual phonons between equal-spin fermions.\footnote{We include BCS among Fermi-liquid theories even if originally it was formulated on a solid atomic crystal, the reason being that modern condensed matter physics, both experimental and theoretical, has been able to reproduce superconductivity in a variety of media, including liquid states.} Fermions within a pair in level $a$ have opposite momentum ${\bf k}_a$. The operator ordering in Eq.~\Eq{bsch} forbids self-interactions \cite{mah}.

It is convenient to introduce the operators
\be\label{bb}
b_a=c_{a-}c_{a+}\,,\qquad b_a^\dagger=c_{a+}^\dagger c_{a-}^\dagger,\qquad N_a=b_a^\dagger b_a,
\ee
where the last operator is the occupation number of level $a$. Equation \Eq{bsch} can be recast as
\ba
H_{\rm BCS} &=&\sum_{a=1}^\Omega H_{\rm BCS}^{(a)}\,,\\
H_{\rm BCS}^{(a)}&=&2\cE_a N_a-\sum_{a'<a}V_{aa'}b^\dagger_{a}b_{a'}\,.\label{bsch2}
\ea
The hard-core $b$ operators satisfy the (anti)commutation relations:
\ba
&&[b_a,b_{a'}]=0\,,\qquad\qquad\qquad [b_a,b_{a'}^\dagger]=\delta_{aa'}(1-2N_a)\,,\label{b1}\\
&&\{b_a,b_{a'}\}=2b_ab_{a'}(1-\delta_{aa'})\,,\qquad \{b_a,b_{a'}^\dagger\}=\delta_{aa'}(1-2N_a)+2b_{a'}^\dagger b_a\,.\label{b2}
\ea
The first one is the same as for bosonic operators, while the others encode Pauli exclusion principle; in particular, on the same energy level $a=a'$ the $b$ algebra is fermionic, $\{b_a,b_a\}=0=\{b_a^\dagger,b_a^\dagger\}$,  $\{b_a,b_a^\dagger\}=1$. Then, for a pair at the $a$-th site, one can define the coherent state $|\Psi\rangle_a=\rme^{\beta_a b_a^\dagger}|0,0\rangle_a=(1+\beta_a b_a^\dagger)|0,0\rangle_a$, where $|0,0\rangle_a$ is the $a$-th copy of the two-particle vacuum state $|0\rangle\otimes|0\rangle$ and $\beta_a$ is a funcion. This suggests to consider a nonperturbative state in the grand canonical ensemble, that is, as a product of individual modes, each corresponding to a paired state occupied by two fermions with opposite spin and momenta:
\be\label{wave0}
|{\rm BCS}\rangle = \exp\left(\sum_{a=1}^{\Omega}\frac{v_a}{u_a} b_a^\dagger\right)|0\rangle_N\,,
\ee
where $|0\rangle_N=\otimes_{n=1}^N |0,0\rangle_a$ is the $N$-pairs vacuum and $u_a$ and $v_a$ are site-dependent functions. As a matter of fact, Eq.~\Eq{wave0} is equivalent to
\be\label{wave}
|{\rm BCS}\rangle = \prod_{a=1}^{\Omega}(u_a+v_ab_a^\dagger)|0\rangle_N\,,
\ee
which we will use from now on. $u_a$ and $v_a$ define a Bogoliubov transformation preserving the algebra if $|u_a|^2+|v_a|^2=1$ (symmetric electrons-holes distribution). Without loss of generality one can take them to be real and parametrized as trigonometric functions: $u_a=\cos\a_a$, $v_a=\sin\a_a$.

The energy of the BCS state Eq.~\Eq{wave} is lower (for every excited electron, by an amount $\Delta$ called \emph{mass gap}) than the free-fermion perturbative vacuum at zero temperature, $(u_a,v_a)=(1,0)$. To show this, one takes the expectation value of the Hamiltonian on the BCS state. The energy relative to the Fermi energy $E_{\rm F}$ (the maximum energy level in the Fermi sea) is
\ba
E_0-E_{\rm F} &=& \sum_a \langle {\rm BCS}|H_{\rm BCS}^{(a)}|{\rm BCS}\rangle\nonumber\\
&=& \sum_a(2\cE_a v_a^2-u_av_a\Delta_a),\label{ene}
\ea
where
\be
\Delta_a \equiv \sum_{a'<a}V_{aa'}u_{a'}v_{a'}\label{delk}\,.
\ee
Minimizing the energy with respect to the probability of the $a$-th site to be occupied, one gets
\be
\frac{\delta E_0}{\delta v_a^2} = 2\cE_a-\Delta_a\frac{u^2_a-v_a^2}{u_av_a}
= 2(\cE_a-\Delta_a\cot 2\a_a)=0\,,\label{zero}
\ee
giving
\be\label{sico}
\sin 2\a_a=\frac{\Delta_a}{\sqrt{\cE_a^2+\Delta_a^2}},\qquad \cos 2\a_a=\frac{\cE_a}{\sqrt{\cE_a^2+\Delta_a^2}},
\ee
where the positive signs are chosen compatibly with the free Fermi sea limit ($\a_a\to 0$). Then
\be
\Delta_a = \frac12\sum_{a'<a}V_{aa'}\frac{\Delta_{a'}}{\sqrt{\cE_{a'}^2+\Delta_{a'}^2}}\,,\label{del}
\ee
and
\be\label{sico2}
u_a=\sqrt{\frac12\left(1+\frac{\cE_a}{\sqrt{\cE_a^2+\Delta_a^2}}\right)},\qquad v_a=\sqrt{\frac12\left(1-\frac{\cE_a}{\sqrt{\cE_a^2+\Delta_a^2}}\right)}.
\ee
Now one can make a mean-field approximation where anisotropic interactions are neglected: the potential is replaced by its average over momenta, which is equivalent to assume that the probability of a state with momentum ${\bf k}$ to be occupied depends only on energy. Therefore $V_{aa'}\to g$ and the order parameter $\Delta_a\to\Delta$ becomes a constant, representing a quasi-particle excitation of energy spectrum $E_a=\sqrt{\cE_a^2+\Delta^2}$. Let $D=2N/(\hbar\omega)$ be the number of Bloch states per unit frequency at the Fermi surface. Taking the continuum limit of Eq.~\Eq{del} (large number of energy levels/lattice sites) and considering an energy transition of a fermionic pair lower than the average phonon energy, $|\cE|<\hbar\omega$, the coupling $g>0$ is
\be
\frac{1}{Dg}=\int_0^{\hbar\omega}\frac{\rmd\cE}{\sqrt{\cE^2+\Delta^2}}\,,
\ee
yielding
\be
\Delta=\frac{\hbar\omega}{\sinh[1/(Dg)]}\,.
\ee
Plugging Eqs.~\Eq{delk}, \Eq{sico}, and $v^2(-\cE)=u^2(\cE)=1-v^2(\cE)$ back into Eq.~\Eq{ene} in the continuum limit, the ground-state energy turns out to be smaller than the Fermi energy, as
\ba
E_0-E_{\rm F} &=& D\int_0^{\hbar\omega} \rmd\cE \{2\cE[2v^2(\cE)-1]-u(\cE)v(\cE)\Delta\}\nonumber\\
&=& D\int_0^{\hbar\omega}\rmd\cE \left(2\cE-\frac{2\cE^2+\Delta^2}{\sqrt{\cE^2+\Delta^2}}\right)\nonumber\\
&=&-\frac{2D(\hbar\omega)^2}{\rme^{2/(Dg)}-1}\nonumber\\
&=&- 2N_*\Delta<0\,,\label{gap}
\ea
where in the last equality we have defined $N_*=N \rme^{-1/(Dg)}$, the number of pairs virtually excited above the Fermi sea ($\Omega\geq N_*+N$; a popular choice is $\Omega=2N$, so that the Fermi sea occupies half of the total available levels). In the weak-coupling limit $Dg\ll1$ ($\Delta\sim 2\hbar\omega N_*/N$, $N_*\to0$), Cooper pairs decorrelate into a free-electron Fermi sea with energy $E_0\approx E_{\rm F}-4\hbar\omega N_*^2/N$, while the strong-coupling regime $Dg\gg 1$ ($\Delta\sim 2Ng$, $N_*\to N$) describes a superconducting medium with energy $E_0\approx -2N\Delta\ll E_{\rm F}$ (average interaction approximation). As the potential between correlated Cooper pairs is attractive, the BCS state is also called a \emph{condensate}. 

Equation \Eq{wave}, which we used to illustrate the mass gap mechanism, is a solution of the Hamiltonian \Eq{bsch2} only in the limit $N\to\infty$. If the number of particles is fixed, then one should consider a coherent superposition of $N$ states. An exact solution of this type was found by Richardson \cite{ric1,ric2} (see also \cite{VB}) and reads, up to a normalization constant,
\ba
|N\rangle_{\rm R} &\propto& \prod_{n=1}^N \sum_{a=1}^{\Omega} \frac{b_a^\dagger}{2\cE_a-e_n}|0\rangle_{N}\\
&\propto& \sum_{a_1,\dots, a_N}\phi_{\rm R}(a_1,\dots, a_N)\prod_{n=1}^N b_{a_n}^\dagger|0\rangle_{N}\,,\label{rs2}
\ea
where $e_n$ are complex solutions of the algebraic equations
\be\label{alg}
\frac{1}{g}+\sum_{n\neq n'=1}^N\frac{2}{e_n'-e_n}=\sum_{a=1}^{\Omega}\frac{1}{2\cE_a-e_n}\,,\qquad n=1,\dots N\,,
\ee
and the energy eigenvalue relative to $|N\rangle_{\rm R}$ is
\be
\cE_{\rm R}=\sum_{n=1}^N e_n;
\ee
the ground state is given by the solutions of Eq.~\Eq{alg} which minimize $\cE_{\rm R}$. In Eq.~\Eq{rs2} the sum excludes more than one pair per level, $\phi_{\rm R}$ is Richardson's wavefunction
\be\label{rich}
\phi_{\rm R}(a_1,\dots, a_N)=\sum_{P}\prod_{n'=1}^N\frac{1}{2\cE_{a_{n'}}-e_{P(n')}}\,,
\ee
and $P(n')$ are the permutations of $1,\dots,n'$.


\section{Quantum spin networks and Fermi-liquid theory}

We are in a position of consolidating the physical picture of the deformed system in the language of framed (or $q$-deformed, or quantum) spin networks \cite{cr91a,cr91b,MSm} (see also \cite{KL,KLo2,KLo1} and the informal description in \cite{smo02}). These structures survive in the degenerate sector, so it may be useful to briefly review them first in the undeformed, non-degenerate case.

The Kauffman bracket is related to Gauss linking number, which is a double integral in three dimensions. This integral diverges when the two integration variables approach each other and the self-linking number requires a point-splitting
regularization \cite{wit89} which ``thickens'' the loops in $\Gamma$ to ribbons. Therefore the natural structure of the Chern--Simons state in loop representation is that of a quantum spin network $\Gamma$, which is a ``fat'' version of spin-network states of quantum gravity \cite{pen71,mou83,RS3}. Edges become two-dimensional tubular surfaces carrying an irreducible representation of the $q$-deformed $su(2)$ algebra of level $k$, denoted as $su(2)_k$ (the parameter $q$ is $q=\rme^{\frac{2\pi \rmi}{2-\rmi k}}$; in the Euclidean case $\rmi k\to k$, to which we stick in the remainder of this section, $q$ is a root of unity). Vertices are promoted to punctured two-spheres, where punctures (the analogue of intertwiners) are the sites where edges connect and are described by conformal blocks of a $SU(2)_k$ Wess--Zumino--Witten (WZW) conformal field theory \cite{WZ,W}.


\subsection{Quantum algebras and WZW model}\label{fsn}

An affine Lie algebra $\mathfrak{g}_k$ of level $k$ is defined by a set of holomorphic currents $J^i(z)$ on the complex plane, $i=1,\dots,\dim\mathfrak{g}_k$, whose operator product expansion (OPE, see e.g. \cite{pol98}) is, up to terms regular as $z_{12}\equiv z_1-z_2\to0$,
\be\label{a1}
J^i(z_1)J^j(z_2)=\frac{1}{z_{12}}f^{ij}_{\phantom{ij}l}J^l(z_2)+\frac{k}{2}\frac{\eta^{ij}}{z_{12}^2},
\ee
where $\eta^{ij}$ is the Killing form and $f^{ij}_{\phantom{ij}l}$ are the structure constants of $\mathfrak{g}_0$ \cite{kac}. In the case of a quantum spin network, the level of the algebra depends on the cosmological constant, Eq.~\Eq{k}. Expanding the generators in Laurent series $J^i(z)=\sum_{m=-\infty}^{\infty}J^i_m z^{-m-1}$, where $J^i_m=(2\pi \rmi)^{-1}\oint \rmd z z^{m}J^i(z)$, Eq.~\Eq{a1} is equivalent to 
\be\label{taun}
[J^i_n,J^j_m]=f^{ij}_{\phantom{ij}k}J^k_{n+m}+\frac{k}{2}\eta^{ij}n\delta_{n+m,0}.
\ee
In particular, the zero modes
\be\label{zerom}
J^i_0=\oint \frac{\rmd z}{2\pi \rmi} J^i(z)=\tau^i
\ee
obey the classical level-0 algebra $\mathfrak{g}_0$. Conformal symmetry is realized by the Sugawara energy-momentum operator
\be\label{sug}
T(z)\equiv \frac{1}{k-h^\vee}\eta_{ij}:J^iJ^j:\equiv\sum_{n=-\infty}^{+\infty}\frac{L_n}{z^{n+2}}\,,
\ee
where $:\,:$ denotes normal ordering,
\be\label{conv}
L_n=\frac{1}{2(k-h^\vee)}\sum_{m=-\infty}^{+\infty} \eta_{ij}:J^i_mJ^j_{n-m}:
\ee
is a Virasoro operator, and $h^\vee$ is the dual Coxeter number, the quadratic Casimir invariant $C_2\mathbb{I}=-\eta_{ij}\tau^i\tau^j$ in the adjoint representation,
\be
h^\vee\delta^l_k=f_{ijk}f^{ijl}\,.
\ee
The OPE 
\be
T(z_1)T(z_2)= \frac{c}{2z_{12}^4}+\frac{2T(z_2)}{z_{12}^2}+\frac{\p T(z_2)}{z_{12}}
\ee
determines a Virasoro algebra 
\be\label{vira}
\left[L_n,L_m\right] = (n-m)L_{n+m}+\frac{c}{12}(n^3-n)\delta_{n+m,0}\,,
\ee
with central charge
\be
c=\frac{k\dim\mathfrak{g}_k}{k-h^\vee}\,.
\ee
For a compact group such as $SU(2)$, unitarity restricts the allowed values of $k$ to be negative integers (rational CFT); in particular, the critical level $k=h^\vee$ is not admitted and Eq.~\Eq{sug} is well-defined. For noncompact groups (an example we shall deal with soon is the special linear group $SL(2,\mathbb{R})$) unitarity requires a finite positive central charge, $k-h^\vee>0$ \cite{KK}.

Specializing to $su(2)_k$ algebra, $\dim\mathfrak{g}_k=3$, $f^{ij}_{\phantom{ij}l}=\e^{ij}_{\phantom{ij}l}$, $\eta_{ij}=\delta_{ij}$, $h^\vee=2$, and $c=3k/(k-2)$. In the fundamental representation, $\tau^i=\sigma^i/(2\rmi)$, where $\sigma^i$ are Pauli matrices. One can complexify the algebra to complex linear combinations of the generators and choose $t^0=\rmi\tau^3,t^\pm=\tau^1\mp \rmi\tau^2$, so that 
\be\label{f}
f^{+-}_{\phantom{+-}0}=2,\qquad f^{0\pm}_{\phantom{0\pm}\pm}=\mp1\,,
\ee
the light-cone metric is
\be\label{eta}
\eta^{00}=-1,\qquad\eta^{+-}=\eta^{-+}=2\,,
\ee
and the quadratic Casimir reads $C_2\mathbb{I}=-\eta_{ij}t^it^j=t^0t^0-\tfrac12(t^+t^-+t^-t^+)$.

In the Wakimoto representation, the algebra is expressed as a free-field $\beta\g$ CFT with a spacelike boson $\varphi$ \cite{wak86,ge90}:
\ba
J^0 &=& \sqrt{\frac{k-2}2}\,\p\varphi+\beta\g\,,\nonumber\\
J^+ &=& \beta\,,\label{pm0}\\
J^- &=& \beta\g^2+\sqrt{2(k-2)}\g\p\varphi+k\p\g\,,\nonumber
\ea
where we have neglected regular terms and $\beta$, $\g$, and $\varphi$ are two commuting ghosts and one boson with respective conformal weight 1, 0 and 0, and OPEs
\ba
\beta(z_1)\gamma(z_2)&=&-\gamma(z_1)\beta(z_2)=\frac{1}{z_{12}}\,,\\
\varphi(z_1)\varphi(z_2)&=&-\ln z_{12}\,.
\ea
To check that Eq.~\Eq{pm0} satisfies Eqs.~\Eq{a1}, \Eq{f}, and \Eq{eta}, one makes use of the above propagators for every contraction between composite operators, and finally expands fields in $z_1$ as, e.g., $\beta(z_1)=\beta(z_2)+z_{12}\p\beta(z_2)+\dots$.

A \emph{chiral} (or \emph{primary}) field $\Phi(z)$ with conformal weight $\Delta_\Phi$ is a holomorphic function such that $L_n\Phi=-z^{n+1}\p\Phi-(n+1)\Delta_\Phi z^n\Phi$; in particular, the Virasoro operator $L_{-1}$ acts on it as a derivative. In the WZW model, primary fields $\Phi_m^j$ are labelled by the total spin $j$ and third component $m=-j,\dots,j$:
\be
\Phi_m^j(z)=\g^{j-m}(z) V_{\a_j}(z)\,,
\ee
where $\a_j=-2\a_0j$ depends on the ``vacuum background charge'' 
\be
\a_0\equiv\pm\frac{1}{\sqrt{2(2-k)}}\,,
\ee
and
\be
V_{\a_j}(z)\equiv  \rme^{\rmi\a_j\varphi(z)}
\ee
is a vertex operator with conformal weight $\Delta_j=\a_j(\a_j-2\a_0)/2$ \cite{DF1}. In a quantum spin network, primary fields live on the tubular edges. $\Omega+1$ such tubular edges carrying different $j$-representations can meet at a two-dimensional node where representations are changed one into another by a multilinear map called intertwiner (or invariant). The change of representation can be thought as a ``scattering'' process from the spin configuration $|j_1,m_1\rangle\otimes\dots\otimes|j_\Omega,m_\Omega\rangle\cong \Phi^{j_1}_{m_1}(z_1)\cdots\Phi^{j_\Omega}_{m_\Omega}(z_\Omega)$ to the outbound state $|j_{\Omega+1},m_{\Omega+1}\rangle\cong\tilde\Phi_{m_{\Omega+1}}^{j_{\Omega+1}}(z_{\Omega+1})$, where $\cong$ denotes the state-vertex isomorphism \cite{pol98} and $\tilde\Phi_{m_{\Omega+1}}^{j_{\Omega+1}}$ is conjugate (in the representation) to the primary field $\Phi_{m_{\Omega+1}}^{j_{\Omega+1}}$ (hence it has the same weight). When $m_{\Omega+1}=j_{\Omega+1}$ is has the form
\be\label{con}
\tilde\Phi_{j}^{j}(z)=\beta^{k-1+2j}(z)V_{2\a_0(k-1+j)}(z)\,.
\ee
Intertwiners are decribed by correlators of primary fields called \emph{conformal blocks} \cite{DF1,DF2,dot90,dot91}:
\be\label{cb}
\phi_\textsc{wzw}(z_1,\dots,z_\Omega)= \left\langle \prod_{a=1}^\Omega \Phi_{m_a}^{j_a}(z_a)\tilde\Phi_{m_{\Omega+1}}^{j_{\Omega+1}}(\infty)\prod_{n=1}^N Q_n\right\rangle,
\ee
where angular brackets are the normalized expectation value in the $\b\g\varphi$ path integral and
\be\label{Q}
Q_n\equiv\oint_{C_n} \frac{\rmd z_n}{2\pi \rmi} \b(z_n)V_{2\a_0}(z_n)\,,\qquad n=1,\dots,N\,,
\ee
is the \emph{screening charge} operator on a contour $C_n$ on the $n$-th edge. Screening operators do not change the conformal properties of the correlators (they have weight 0 and are representations of the algebraic identity) and contribute to the charge neutrality conditions on the conformal block. These are determined by  Eq.~\Eq{con} with $j=0$, the conjugate to the identity operator, and read \cite{dot90}
\be\label{cnc}
j_{\Omega+1}=\sum_{a=1}^\Omega j_a-N\,,\qquad m_{\Omega+1}=\sum_{a=1}^\Omega m_a\,.
\ee
The complex coordinate system in Eq.~\Eq{cb} was chosen to place the bosonic vacuum charge $2\a_0$ at the pole of the conformal sphere ($z_{\Omega+1}=\infty$), away from the charges in the $\b\g$ sector.

Applied to a highest weight null vector $|v\rangle$, Eq.~\Eq{conv} with $n=-1$ reads
\ba
0&=&\left[L_{-1}-\frac{1}{2(k-h^\vee)}\sum_{m=-\infty}^{+\infty} \eta_{ij}:J^i_mJ^j_{-1-m}:\right]|v\rangle\nonumber\\
&=&\left(L_{-1}-\frac{1}{k-h^\vee} \eta_{ij}J^i_{-1}J^j_0\right)|v\rangle.
\ea
The state-operator isomorphism then yields the Knizhnik--Zamolodchikov equations for the conformal block \Eq{cb} \cite{KZ}:
\be\label{kz}
\left[(k-h^\vee)\p_a+\sum_{a'\neq a}\frac{\eta_{ij}\tau^i_{(a)}\tau^j_{(a')}}{z_a-z_{a'}}\right]\phi_\textsc{wzw}(z_1,\dots,z_\Omega)=0,\qquad a=1,\dots,\Omega\,,
\ee
where $\tau^i_{(a)}$ is a generator of the $su(2)$ algebra in representation $j_a$ acting on the corresponding primary field.

The space of conformal blocks with different choices of representations, number of edges $\Omega$, number of charges $N$, and contours $C_n$ enters the definition of the space of framed spin networks.


\subsection{BCS quantum gravity}\label{inte}

The connection between BCS theory and degenerate LQG can be better formalized by recognizing the former as a deformed CFT, namely, an $SL(2,\mathbb{R})_{k=2}$ WZW model at critical level \cite{sie99} (see also \cite{AFS}).\footnote{As  Chern--Simons theory in a three-dimensional space induces a two-dimensional WZW model on the boundary \cite{wit89,MS,EMSS}, Eq.~\Eq{eom3} already hints at this relation.} The $SU(2)_k$ WZW theory was reviewed in the previous subsection. Although unitarity of compact groups excludes the critical level $k=h^\vee$, the complexification of the $su(2)_k$ algebra Eq.~\Eq{f} coincides with that of the real algebra $sl(2,\mathbb{R})_k$ of $2\times2$ matrices with vanishing trace; this is a subalgebra of the Virasoro algebra \Eq{vira} with generators $L_0$, $L_\pm$. The corresponding special linear group $SL(2,\mathbb{R})_k$ of $2\times2$ matrices with determinant 1 is  noncompact and admits a well-defined critical limit, where the field $\varphi$ with infinite bakground $U(1)$ charge $\a_0$ decouples from the spectrum. In fact, the rescaled Virasoro operators $l_n\equiv (k-h^\vee)L_n$ obey an abelian algebra ($[l_n,l_m]=0$), as one can see by taking the critical level in Eq.~\Eq{vira}.\footnote{The symmetries of WZW models for $k=2$ were studied in \cite{FF,BK2,BS1,BS2}. The $W$-symmetry \cite{BoS} of the coset $SL(2,\mathbb{R})_k/U(1)$ WZW model \cite{wit91,KKK}, which shares many features of the ungauged $SL(2,\mathbb{R})_k$ model, is described by a deformation of the linear $W_{\infty}$ algebra, dubbed $\hat W_{\infty}(k)$, where the level $k$ is responsible for nonlinearities \cite{BK2,BS1,BS2}. (A $W_{\infty}$ algebra is a linear extension of the Virasoro algebra to generators $L^j(z)=\sum_mL_m^jz^{-m-(j+2)}$ with conformal spin $j+2$, $j\in\mathbb{N}$ \cite{PRS1,PRS2}. A sector of classical self-dual gravity, which may have some relation with the present framework, is parametrized by elements of this algebra \cite{pa90a,pa90b,hus92}.) The spectrum of $W_\infty$ and $\hat W_{\infty}(k)$ is the same and the generators are chiral fields with integer spin $j\geq 2$. When $k\to\infty$, $\hat W(k)$ reduces to the standard $W_\infty$ algebra, while in the critical limit $k\to 2$ $\hat W(k)$ is truncated to primary fields with integer spin $j\geq 3$; $\hat W_{\infty}(2)$ is linear and can be written in closed form \cite{BS1,BS2}. 
}

The BCS model features $\Omega-2N$ empty levels with spin $m_a=1/2$, while all levels are in the fundamental representation ($j_a=1/2$ for all $a$). Therefore the charge neutrality conditions Eq.~\Eq{cnc} reduce to
\be
j_{\Omega+1}=m_{\Omega+1}=\frac{\Omega}2-N\,.
\ee
Energy levels are promoted to complex coordinates $z_a=2\cE_a$ and Richardson's wavefunction, Eq.~\Eq{rich}, to a function on $\mathbb{C}^\Omega$, $\phi_R(a_1,\dots, a_N)\to\phi_R(z_1,\dots, z_\Omega)$. Introducing the nonconformal operator
\be\label{vo}
V_0\equiv \exp\left[-\frac{\rmi\a_0}{g}\oint_{C_0} \rmd z\, z\p\varphi(z)\right],
\ee
one can build a deformed block $\phi_\textsc{wzw}^*(z_1,\dots,z_\Omega)\equiv\langle V_0\dots\rangle$, where dots indicate the same insertions as in Eq.~\Eq{cb} (all primary fields with equal spin $j_a=1/2$) and the contour $C_0$ encircles all other coordinates $z_a$. As showed in \cite{sie99}, to which we refer for an extensive proof, this object is proportional to Richardson's wavefunction in the limit where the Liouville field $\varphi$ decouples from the system ($k\to 2^+$, $\alpha_0\to\infty$). More precisely,
\be
\phi_{\rm R}(z_1,\dots,z_\Omega)\ \stackrel{\a_0^2\to -\infty}{\sim}\  (-2\sqrt{\pi}\a_0)^N\sqrt{\det A}\,\rme^{\a_0^2 U}\,\phi^*_\textsc{wzw}(z_1,\dots,z_\Omega)\,,
\ee
where
\be
U=\sum_{a}^\Omega\sum_{n}^N \ln z_{an}^2-\sum_{a<a'}^\Omega \ln z_{aa'}-\sum_{n<n'}^N\ln z_{nn'}^4+\frac1g\left(2\sum_n^N z_n-\sum_a^\Omega z_a\right)\,,
\ee
and $A=-\p_n\p_{n'}U/2$. The limit to critical level is responsible for setting, on each site $n=1,\dots,N$, the point $z_n\equiv e_n$ inside the contour $C_n$ defining the $n$-th screening charge. The relation between Richardson's wavefunction and the conformal block \Eq{cb} is
\be
\phi_{\rm R}(z_1,\dots,z_\Omega)\ \stackrel{\a_0^2\to -\infty}{\sim}\  \exp\left[\a_0^2U+\frac{\a_0^2}{g}\left(\sum_a^\Omega z_a-2\sum_n^N z_n\right)\right]\phi_\textsc{wzw}(z_1,\dots,z_\Omega)\,.
\ee

The number of eigenstates of $H_{\rm BCS}$ equals the number ${\Omega\choose N}$ of possible contour choices in a block, so Hamiltonian BCS eigenstates are in correspondence with deformed conformal blocks. Moreover, as in the BCS theory there are $\Omega$ copies of the algebra Eq.~\Eq{f} with generators $t^0_{(a)}=N_a-1/2$, $t_{(a)}^\dagger=-\rmi b_a$, and $t^-_{(a)}=-\rmi b_a^\dagger$, it is natural to identify (compare Eq.~\Eq{zerom}) the hard-core creation-annihilation operators \Eq{bb} with the algebra zero modes on the $a$-th edge:
\bs\label{bj-}\ba
b_a				  &\leftrightarrow& \rmi\oint \frac{\rmd z_a}{2\pi \rmi} J^+(z_a)\,,\\
b_a^\dagger &\leftrightarrow& \rmi\oint \frac{\rmd z_a}{2\pi \rmi} J^-(z_a)\,,\\
N_a         &\leftrightarrow& \frac12+\oint \frac{\rmd z_a}{2\pi \rmi} J^0(z_a)\,.
\ea\es
Then one can argue that the Hamiltonian $H_{\rm BCS}^{(a)}$ acts, up to constants and a linear term in $z_a$, as $l_0^{(a)}$ on a primary state. This result can be obtained also by comparing the integrals of motion of the BCS theory \cite{CRS} with the Knizhnik--Zamolodchikov equations \Eq{kz} \cite{sie99}.

To summarize, BCS levels are mapped to edges of quantum spin networks, interacting at nodes via a BCS coupling. On $N$ out of $\Omega$ edges there lives a copy of the identity element represented by a screening charge $Q_n$ and corresponding to a Cooper pair. The interaction among pairs is given by an operator breaking conformal invariance and evaluated along a loop encircling all the pairs. This contour is a loop on a two-dimensional node whose interior includes all the $\Omega$ punctures and $N$ screening charges. Conformal invariance and the undeformed WZW model are recovered in the strong coupling limit $g\to\infty$ (large gauge-field values).

The BCS interaction is the cornerstone of the construction of three-dimensional geometry from quantum spin networks. Classically, it describes scattering of Jacobson electring lines at their endpoints; these worldsheets, patched together at their edges, span the three-dimensional spacetime, giving rise to the geometric sector which was lost in the free-field picture. Before quantizing the theory, we fixed the gravitational configuration to be a particular degenerate sector. This way of proceeding looks similar to that of loop quantum cosmology and minisuperspace models, where a reduction of the symplectic structure is performed at classical level. However, in our case that reduction is dictated by a dynamical cosmological constant and is not optional. What is arbitrary, as far as we can see, is the choice of the degenerate sector between the two cases ${\rm rk} E=1$ and ${\rm rk} E=2$; we concentrated on the second, namely Jacobson's configuration. Furthermore, selection of a degenerate sector does not result in a loss of spacetime dimensionality, as we have just argued.

The screening charges at a given node have an intuitive picture as the sites activated in an area measurement, i.e., when a classical area intersects the spin network. In condensed matter physics, the Fermi sea is a ground state of uncorrelated electron pairs whose Fermi energy is higher that the BCS pair-correlated state. In quantum gravity, pair correlation can be regarded as a process of quantum decoherence. An abstract spin network is the gravity counterpart of an unexcited Fermi sea. As soon as an area measurement is performed on the state, $N$ edges of a given node are activated and the system relaxes to a lower-energy vacuum corresponding to the selection of one of the area eigenstates in a wave superposition. In a sense, physical areas are Fermi-liquid surfaces and measuring geometry equals to counting Cooper pairs.

Given these and other forms of evidence below, we posit the following
\begin{quote}
{\bf Conjecture (version 1):} \emph{Loop quantum gravity with a cosmological constant is dual to a two-dimensional Fermi liquid living on an embedded spin network}.
\end{quote}
This correspondence was showed in the particular case where all spin labels in the network are in the fundamental representation. This is a consequence of the assumption, made in the BCS theory, that energy levels can host only one Cooper pair. If one lifted this degeneracy by allowing $d_i$ pairs per level, the associated primary fields would have spin $d_i/2$.

A further generalization to arbitrary spin networks and non-BCS Fermi-liquid theories pertains the choice of the algebra level. Quantum spin networks carrying irreducible representations of a noncritical algebra are possibly described by non-mean field theories. In the large $N$ limit the outbound WZW state $\langle\phi^*_{\rm WZW}|\sim\langle0|\rme^Q$ ($Q$ is a screening charge) resembles the coherent state Eq.~\Eq{wave0} in the grand canonical ensemble \cite{sie99,DF1}, allowing the correspondence $\sum_a(v_a/u_a)b_a\leftrightarrow Q$. Equations \Eq{pm0}, \Eq{Q}, and \Eq{bj-} lead to the identification \cite{sie99} of the phase $\vartheta$ of the order parameter $v_a/u_a$ with $\a_0\varphi(z)$. For large $\a_0$ (large phase stiffness), one recovers the BCS model where the value of the critical temperature does not depend on $\vartheta$. On the other hand, a finite $\a_0$ leads to non-mean field theories where the phase of the order parameter is dynamical; examples in condensed matter physics are discussed in \cite{DI,EK1,EK2}.

The BCS theory is a (deformed) CFT living on a $(1+1)$-dimensional submanifold of a $3+1$ classical spacetime described, at the quantum level, by the Chern--Simons state. Despite the notable differences with respect to the usual AdS/CFT holography (the ``boundary'' here is of codimension 2 and the field theory on it is not conformal everywhere),
our model well shares the main features of the correspondence recently formulated in \cite{fre08} in the context of quantum gravity: namely, rather than asking what is the CFT at the boundary of a given spacetime (here, de Sitter), one is able to associate a given two-dimensional theory (in this case, a deformed WZW model) for a given vacuum state (here, the Chern--Simons state).

We can make this statement more precise and show that the dS entropy emerges from microscopic degrees of freedom which are nothing but wormholes opening and closing on the boundary. These wormholes are the geometric equivalent of Cooper pairs. Therefore another formulation of the above conjecture is
\begin{quote}
{\bf Conjecture (version 2):} \emph{Loop quantum gravity with a cosmological constant is dual to a two-dimensional Fermi liquid living on the time-space boundary of a de Sitter background}.
\end{quote}

We provide an explicit example of quantum gravity of de Sitter space with fermions that propagate on the cosmological horizon. The de Sitter bulk solution becomes a monopole in Ashtekar formulation. The magnetic field $B^i_\a$ is sourced by a monopole which arises from mappings, in the connection representation, from target space to the internal space. This classical solution was found in \cite{Kam91}. The de Sitter metric in static coordinates is
\be 
\rmd s^2=-\left(\frac{12-\Lambda r^2}{12+\Lambda r^2}\right)^2\rmd t^2+\left(1+\frac{\Lambda}{12}r^2\right)^{-2}(\rmd r^2+r^2\rmd\Omega_2^2),
\ee
where $r$ is a radial coordinate and $\Omega_2$ is the line element of a 2-sphere (with angular coordinates $\varphi$ and $\vartheta$). Spatial sections are 3-spheres with radius $\sqrt{3/\Lambda}$. The connection and triad read ($x^\a=r,\varphi,\vartheta$)
\ba
A_{\a}^i &=&  \frac{\Lambda}6\frac{\e^i_{\phantom{i}\a\b} x^\b}{1+\tfrac{\Lambda}{12}r^2}\,, \label{mo1}\\
E_i^\a &=& \frac{\delta^\a_i}{\left(1+\tfrac{\Lambda}{12}r^2\right)^2}\,,
\ea
while the field strength is
\be 
F_{\a\b}^i=\frac{\Lambda}3\frac{\epsilon^i_{\phantom{i}\a\b}}{\left(1+\tfrac{\Lambda}{12}r^2\right)^2}\,.\label{mo3}
\ee
The triad is only $r$ dependent and determines radial fluxes. 

Note that, as $A=O(1/r)$ in the radial direction for large $r$, the holonomy near one of the poles of $S^3$ is $h\approx 1+\int A$ to lowest order; qualitatively, in the $r\to\infty$ limit where perturbative gauge theory is valid, the connection representation is a reliable approximation of holonomy representation.

One can chart the two hemispheres $S^3_\pm$ of $S^3$ with different coordinate systems. The boundary of $S_\pm^3$ is a two-sphere $S^2$ which can be approached by a gauge transformation which fixes radial iso-vectors to the $i=3$ direction. This corresponds to a Dirac string supported by the $U(1)$ monopole $A^3_\a$. The magnetic field associated with this component gives a nonvanishing monopole charge on the horizon. The components $A^{i\neq 3}_\a$ vanish on $S^2$ but not inside the horizon, and they source the magnetic charge density. By stereographic projection, one concludes that there is a monopole on the north pole of $S^3$ and an anti-monopole on the south pole.

\begin{figure}\begin{center}
\includegraphics[width=10cm]{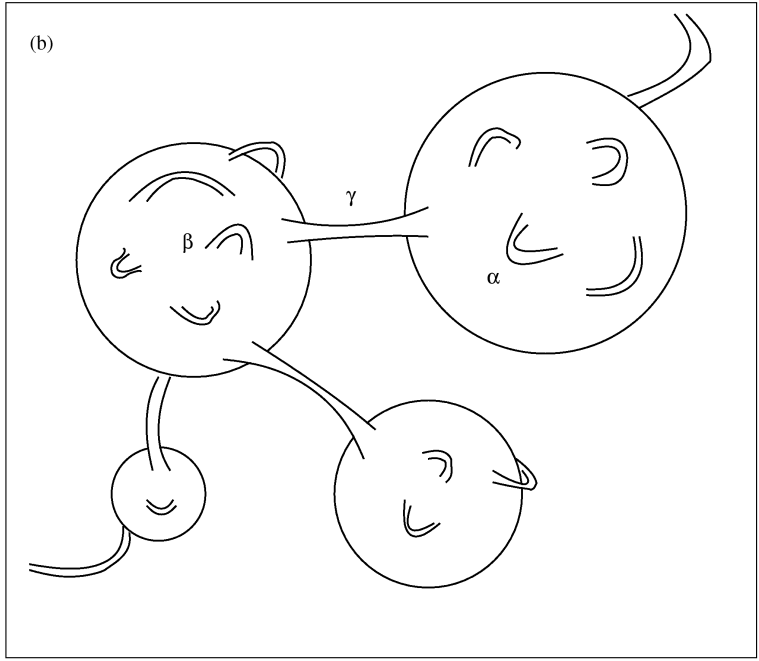}
\caption{\label{fig1} Nonlocal correlations on (tubes $\a$ and $\b$) and between (tube $\g$) dimensionally reduced $S^3$ de Sitter horizons.}
\end{center}\end{figure}

As one can see from Eqs.~\Eq{mo1}--\Eq{mo3}, the monopole--anti-monopole configuration coincides with Jacobson's sector.
Taking an arbitrary number of gravielectric lines, one ends up with the picture of a dS boundary populated by oppositely charged singularities, i.e., wormholes. Their endpoints can be either on the same horizon or connect horizons of different observers (Fig.~\ref{fig1}). The total area of an horizon is given by counting the number of paired punctures. On the other hand, these provide the quantum degrees of freedom which account for thermodynamical properties of gravity, so that the Bekenstein--Hawking law \cite{B,H} relating the horizon area to the dS entropy (see \cite{SSV,Bou02} for reviews) is naturally recovered at leading order and given the usual physical meaning.


\section{Discussion}\label{disc}

Quantum gravity in a four-dimensional causal degenerate sector with a deformed topological structure mimics the behaviour of lower-dimensional Fermi liquids and, in the simplest case, is actually equivalent to a BCS theory of superconductivity. It is known that the nonlocal nature of LQG prevents the formation of what are classically regarded as singularities \cite{Bo06a,Bo06b,Ash08}; roughly speaking, quanta of geometry cannot be compressed too densely and they determine the onset of a repulsive force at Planck scale \cite{Ash08}. Here we have seen that the agent behind this mechanism is literally Pauli exclusion principle, thus realizing in a precise way the long-pursued idea that quantum gravity can exhibit Fermi--Dirac statistics \cite{FS1,FS2,sam93}. The details of the possible multi-fermion interactions in gravitational degenerate sectors will require further attention and the wisdom of condensed matter models.\footnote{For instance, fermion-phonon interactions may be considered and one could be able to describe simple crystal-like spacetime configurations in terms of Bloch models.} 

Here is an example of the possibilities opened up by the Fermi liquid viewpoint. The BCS weakly-coupled phase ($g\ll 1$) describes a Fermi sea of free fields with fermionic statistics. This false vacuum (gravity at short scales) is interpolated to the nonperturbative ground state with Bose statistics (small total occupation number $N=\sum_a N_a$),\footnote{On the true vacuum $|{\rm BCS}\rangle$ the occupation number operator $N_a=0$ for all $a$'s. Then, according to Eqs.~\Eq{b1} and \Eq{b2}, the hard-core operators obey a bosonic algebra.} and a quantum statistics transition takes place. These two regimes find consistent counterparts in the other sectors of the BCS/gravity correspondence. Actually, we have seen that the Chern--Simons state is of the form $\Psi_{\rm CS}\sim \rme^{f(x)Q^5}$, where $f(x)$ is a function of the transverse coordinates, $Q^5\sim \int \rmd z\, n_z j^{5z}$ is the charge associated with the axial current, and $n_z$ is some gauge-fixed vector component proportional to $E_z^i$. The statistics of the Majorana spinors constituting this current is fermionic and in the perturbative limit the cosmological constant assumes its ``bare'' value $\Lambda_0$. On the other hand, the BCS/WZW duality states that BCS theory at strong coupling ($g_{\rm BCS}=g\gg 1$, small $N$, Bose statistics, nonperturbative true vacuum) is mapped into a WZW model where the coupling $g_{\rm WZW}=k-2$ between the boson $\varphi$ and the commuting (i.e., ghost) fermionic system $\b\g$ is weak. The wavefunction of the strongly-coupled system is again of the form $\rme^Q$, where the current now is Eq.~\Eq{bj-} and the spin-statistics has changed to bosonic. Therefore we can postulate that there is a Bose--Einstein condensate (BEC)-BCS crossover from weak to strong coupling, as recently studied in composite-fermion theory \cite{HCJV}.

There are many other features of the gravity/BCS model we have simply hinted at without a complete proof. First, the form of the cross-coupling $V_{kk'}$ in Eq.~\Eq{vkk} is determined by Eqs.~\Eq{fou} and \Eq{mass2} and is still to be computed. In turn, this should clarify the correspondence between the BCS ground state Eq.~\Eq{wave0} and the deformed Chern--Simons state (see Eq.~\Eq{ycspsi}). Third, the energy spectrum of the BCS model should reproduce the LQG spectrum of the area operator \cite{rov04}, corrected by the mass gap. The latter is in fact a nonperturbative redefinition of the cosmological constant,
\be
\langle\Lambda(\psi)\rangle=\Lambda_0 \langle\exp (-j^{5z})\rangle\,,
\ee
which can address the related smallness problem in terms of a condensate with vacuum expectation value $\langle j^{5z}\rangle\sim O(10^2)$. In the perturbative regime (small values of the connection), $\langle\Lambda\rangle\approx\Lambda_0(1-\langle j^{5z}\rangle)=O(1)$. In the nonperturbative regime, the effective mass grows important and the cosmological constant, supposing $\langle j^{5z}\rangle$ to be positive definite for large connection values, becomes exponentially small. Thus the smallness of the cosmological constant is regarded as a large-scale nonperturbative quantum mechanism not dissimilar from quark confinement: quantum gravity at short scales seems to admit a perturbative formulation.

Another interesting datum to store for the future is that critical algebras have been invoked in string-theoretical systems with strong gravitational fields. Therein, strings seem to behave as tensionless objects at short distance \cite{DS1,DN,DGN} (see \cite{sch77,KaL} for classical strings). As the Regge slope $\alpha'\sim (k+h^\vee)^{-1}$, critical algebras describe the tensionless string spectrum \cite{BS1,GM1,GM2,gro88,LZ,sav1,sav2}. Due to the aforementioned contraction of the Virasoro algebra, a critical dimension does not arise \cite{BRSS} and perturbative string methods are inadequate. In this limit, gravity is decoupled from the other fields \cite{BS1}, leading to a sort of no-target configuration.

In Ashtekar formalism there is still a causal structure and a residual geometry, constituted classically by interacting electric lines and, at quantum level, by framed spin networks. These are the sought-for target background associated with tensionless strings. Remarkably, this picture is consistent with the general notions that
higher-spin interactions (as those stemming from $\hat W(k)$ algebras \cite{BS1,BS2}) require a non-vanishing cosmological constant \cite{FV} and, on the other hand, can be described by Chern--Simons gauge theories \cite{ble89,EH}. Furthermore, critical algebras are related to those with large level $k'$ via the Langlands duality  $k'-h^\vee=(k-h^\vee)^{-1}$ \cite{FF}. The limit $k\to\infty$ corresponds to a semiclassical limit in string theory \cite{BS2}. Readily, in loop quantum gravity its counterpart is semiclassical general relativity with vanishing cosmological constant.\footnote{The adjective ``semiclassical'' is suggested by the qualitative idea that the Kodama state is well-defined in the $\Lambda\to 0$ limit if the Chern--Simons form is made to vanish, too; then the connection $A$ is small.}

While in superstring cosmological scenarios a dynamical mechanism is needed to reduce the dimensionality of spacetime from ten to four \cite{BV}, we are faced with a similar charge but in the opposite direction, namely, how to get four dimensions from a lower-dimensional degenerate sector. Given the technology developed above, we believe this problem is actually related to the measurement problem and the decoherence approach (see \cite{Zu03} for a review and, in particular, \cite{sak88,hal89,pad89,kie93,ASc}). In the latter, one can construct generalized decoherence functionals for coherent states, and wavefunctions loose phase correlations due to interaction with the environment. The BCS wavefunction can be regarded as a coherent state describing quantum spin networks; we noticed that the activation of network edges (via identity representations/screening charges) corresponds to area measurements. If these areas were macroscopic, the decoherence functional would be able to yield a macroscopic measurement from the quantum ones via phase correlations. Coherence properties would have a counterpart in the number of screening charges and contour choices in the WZW model.

The emergence of a two-dimensional theory near Planck scale seems to be a robust feature of independent quantum 
gravity models. Apart from string theory, we mention the asymptotic safety scenario \cite{LaR1,LaR2,NiR} and the 
causal dynamical triangulations (CDT) approach \cite{AmJ,AgM,AL1,AJL1,AJL2,AJL3,AJL4,AJL5,AJL6,AGJL} (for a review 
consult \cite{lol08}), both of which predict a fractal lower-dimensional structure at short scales. Notably, in CDT 
quantum gravity the two-dimensional regime, made of a superposition of interacting quanta of geometry, dynamically 
generates a four-dimensional de Sitter spacetime at large scales. Although it is premature to draw comparisons between 
these works and ours, there is an intriguing convergence of findings worthy of future study.

In $2+1$ dimensions, a nonrelativistic Fermi-liquid theory was invoked in relation with a noncritical version of 
M-theory \cite{HoK}. Another $(2+1)$-dimensional BCS model has been studied recently with techniques of the AdS/CFT correspondence \cite{HHH}. Spaces with dimension 2 are important because they allow anyonic statistics which, in turn, may be involved in the spin-statistics transition outlined above. The ${\rm rk} E=2$ degenerate sector may play a role in this respect.

Finally, in contrast with more ambitious frameworks such as string theory, loop quantum gravity does not claim to be a ``theory of everything'', as matter fields are inserted by hand. However, it has been long conjectured (and recently given some evidence \cite{BMS,BHKS}) that matter excitations might be realized as particular states in the spin network space, thus ideally fulfilling Einstein's view of matter and geometry as an unicum. The picture we have here proposed goes towards the same direction: the two-dimensional Fermi-liquid (in particular, BCS) theory naturally reproduces the structure of framed spin networks, where braid configurations (corresponding to standard model generations) are expected to live. Stated in a slightly different language, the Fermi-liquid Hamiltonian may be regarded as the holographic generator of a four-dimensional fermionic sector, thus advancing the possibility to regard matter as an epiphenomenon in a symmetry-broken mother theory. Scalar and vector particles, on the other hand, might be obtained via bosonic condensation in the Fermi liquid \cite{FGT}. Needless to say, all these speculations have yet to be proven.


\ack

We thank A. Ashtekar, A. Corichi, L. Freidel, R. Gambini, E.E. Ita, T. Jacobson, J.K. Jain, S. Mercuri, J. Pullin, A. Randono, G. Sierra, J. Sofo, D. Vaid, and especially J.D. Bjorken, M. Bojowald and L. Smolin for useful discussions. S.H. is supported by NSF CAREER Award and G.C. by NSF grant PHY-0653127.


\end{document}